\documentclass[10pt,twocolumn,twoside,journal,a4paper]{IEEEtran}

\usepackage{amsmath,amssymb,amstext,mathtools,tikz}
\usetikzlibrary{positioning,calc}
\tikzset{
  wagon/.style = {circle, draw, thick, fill=white, minimum size=18pt, inner sep=0pt, font=\small},
  bond/.style  = {thick},
}
\newcommand{\ttdots}[1]{%
  \foreach \d in {-6pt,0pt,6pt}{\fill ($#1+(\d,0)$) circle (1.1pt);}}
\newcommand{\ttellipse}[5]{%
  \draw[bond] (#1pt,#2pt) ellipse ({#3pt} and {#4pt});
  \node at (#1pt,#2pt) {#5};}
\newcommand{\ttlegs}[8]{%
  \foreach \k in {1,...,#5}{%
    \pgfmathsetmacro\xx{#6+(#7-#6)*(\k-1)/(#5-1)}%
    \pgfmathsetmacro\yy{#2-#4*sqrt(1-((\xx-#1)/#3)^2)}%
    \draw[bond] (\xx pt,\yy pt) -- (\xx pt,#8pt);}}
\usetikzlibrary{shapes.geometric}
\tikzset{idx/.style={inner sep=1pt, font=\footnotesize},
         blab/.style={idx, above=7.5pt},   
         wwagon/.style={ellipse, draw, thick, fill=white, minimum width=24pt,
                        minimum height=18pt, inner sep=0pt, font=\small}}  
\newcommand{\bondlab}[3]{\draw[bond] (#1) -- node[blab]{#3} (#2);}
\newcommand{\legdown}[3]{\draw[bond] (#1) -- (#1 |- 0,#2pt)
   node[idx, right, pos=0.55]{#3};}
\newcommand{\ttp}{\mkern-2mu{+}\mkern-2mu}
\newcommand{\ttm}{\mkern-2mu{-}\mkern-2mu}
\newcommand{\ttstub}[4]{\draw[bond] (#1.#2) -- node[blab]{#4} ++(#3pt,0);}
\tikzset{tallwagon/.style={ellipse, draw, thick, fill=white, minimum width=18pt,
                           minimum height=#1, inner sep=0pt, font=\small}}

\usepackage{graphicx}
\usepackage{color}

\def\comment#1{}

\newcommand {\mymat}[1]  {\mathbf{#1}}

\newcommand {\ba} {\mymat{a}}

\newcommand {\bc} {\mymat{c}}

\newcommand {\bz} {\mymat{z}}

\newcommand {\bI} {\mymat{I}}

\newcommand {\X} {\mymat{X}}

\newcommand {\tT} {{\cal{T}}}

\usepackage{listings}
\begin{document}

\title{Algebraic Operations on Tensor Trains
\thanks{This work was supported by the Czech Science Foundation
through the project No. 25-18070S.}
}

\author{
Petr Tichavsk\'y\\
\textit{Institute of Information Theory and Automation of the
Czech Academy of Sciences} \\ Prague 8, Czech Republic \\
tichavsk@utia.cas.cz}

\maketitle

\begin{abstract}
The tensor train (TT) model is widely used to approximate
high-dimensional tensors, enabling efficient handling of data that
may exceed available memory. TT helps address the curse of
dimensionality in applications such as system identification and
dynamic programming. In some applications, TT is known as a
``matrix product state" (MPS). This paper introduces algorithms
that facilitate (1) approximation of a TT with another TT with a lower bond
dimension (known as tensor rounding)
(2) the summation, (3) Hadamard (elementwise) product, and
(4) matrix--vector product of matrices and vectors (tensors) represented
in the tensor train (TT) format. The last product is also known under
the acronym MPO--MPS. The proposed algorithms achieve an improved
trade-off between computational efficiency and accuracy compared to
state-of-the-art methods.
\end{abstract}



\section{Introduction}

Multidimensional data structures, known as tensors, frequently arise
in applications where their size makes complete storage impractical.
Such large tensors are typically generated by sampling a multivariate
function over a hyper-rectangular grid \cite{ssp2023}. The number of tensor elements
increases exponentially with the number of dimensions, or tensor order.
In quantum chemistry, the matrix product state, also referred to as a
tensor train (TT), is used to represent the wave function of a many-particle
system \cite{white_prl}. In this context, the number of dimensions corresponds to the
number of particles, which can be considerable. Consequently, it is
necessary to perform algebraic operations on these tensors without
evaluating all individual elements.

In addition to the TT \cite{TT,Novikov}, there are other tensor decomposition formats,
including canonical polyadic decomposition (CPD) \cite{parafac,candecomp}, Tucker
decomposition \cite{Tucker}, and the sum of tensor trains (SOTT) \cite{SOTT}. A comparison
of these formats is presented in Section II. We also present methods for
converting one tensor format to another.

A key question is how to approximate a tensor given in TT format by
another tensor in TT format with a lower bond dimension. This problem
arises in several applications, such as parametric low-rank kernel
approximation \cite{LKA}, high-dimensional partial differential equations  \cite{HPDE},
thermal radiation transport \cite{TRT}, and the solution of linear tensor
equations \cite{LTE}. In practice, the bond dimensions of TT models can become
very large. In quantum chemistry, the bond dimensions of the involved
TTs (matrix product states) can easily be 1000 or more.

Reducing the bond dimension is known as tensor rounding (TR).
The standard TT approach to rounding, proposed by Oseledets  \cite{TT}, has
two phases \cite{TT}, Algorithm 2: orthogonalization followed by compression
(typically using singular value decomposition, SVD). Here, by orthogonalization,
we mean a sweep of orthogonalization steps across every tensor core.
Analysis shows that the orthogonalization step dominates the computational
cost of this approach. Several novel algorithms for tensor rounding have been
proposed recently, including the Gram SVD algorithm \cite{Parallel,Ballard}, a class of randomized
TR algorithms \cite{Randomized}, and TR using the Khatri--Rao product (KRP) \cite{Adaptive}.  These
algorithms exhibit improved speed compared to the traditional algorithm.
Although their asymptotic complexity is about the same, they can be up to
50$\times$ faster in practice. Unfortunately, these algorithms appear to be less
accurate if the tensor rounding is not lossless.

There are two possible cases: either the original TT can be accurately
approximated by a TT with a lower bond dimension, or it cannot. In the
former case, KRP-based and randomized TR methods work fine and are fast.
In the latter case, the approximation incurs a loss in accuracy, and the
algorithms exhibit a tradeoff between accuracy and complexity (bond dimension).
Usually, the original TR algorithm offers a better tradeoff between accuracy
and complexity than the novel algorithms; we observe this behavior in our simulations.

Recently, another novel tensor-rounding algorithm was proposed in \cite{Optim}.
The latter algorithm is a variant of a TT--SVD (TTSVD) algorithm \cite{TT}.
Recall that the original TTSVD algorithm builds the TT model for a tensor
in which all elements are accessible. In \cite{Optim}, the TTSVD algorithm was modified
to enable it to work with a tensor already represented in TT format but with a
higher bond dimension. In this paper, we propose a more efficient
implementation of the algorithm, and call it TTSVD-TT0. In our simulations,
we show that the algorithm has a nearly identical tradeoff between accuracy
and complexity as the original TT rounding, but it is somewhat faster.
Moreover, it is more accurate than the novel TT rounding algorithms in lossy
scenarios.

Second, we present another novel variant of the algorithm, which is more computationally efficient than
TTSVD-TT0 in cases when the bond dimension of the outcome is significantly lower than
the bond dimension of the original tensor. We call the latter
algorithm TTSVD-TT1 for easy reference. Unlike TTSVD-TT0, the novel variant performs rounding in two separate steps.

The main contribution of this paper is a further generalization of the
TTSVD-TT1 algorithm. The input tensor might not be given
as a single TT but instead as (a) a sum of several TTs, (b) a Hadamard
(elementwise) product of two TTs, or (c) a matrix--vector product of two TTs,
where one TT represents a matrix and the second TT represents a vector to be
multiplied. Task (c) is also known under the name of the compressed matrix product
operator and matrix product state (MPO--MPS) product \cite{Successive}. The latter algorithm
is called Successive Randomized Compression (SRC). We propose an alternative
approach that might be more accurate, as shown in simulations.

Although many applications are possible, we consider one application where the
tensors come from quantum chemistry \cite{Quantum}. The main task is to solve a constrained
quadratic optimization problem, which can be viewed as a minimum eigenvalue
problem for a large-dimensional Hamiltonian \cite{Inexact}. The Hamiltonian $H$ is an
order-$N$ tensor, where $N$ denotes the number of sites in a molecule. The Hamiltonian
can be defined through its canonical polyadic (CP) decomposition. The CP decomposition
has CP rank of asymptotic order $O(N^4)$. We convert the CP format to TT format.
Then, assume that we have an initial estimate of the wave function $x$ in TT format.
We show how to compute the product $y = Hx$, where $y, x$, and $H$ are all represented
in TT format. With this tool, it is possible to minimize the quadratic form
$x^T H x$ subject to $\|x\| = 1$, i.e., find the wave function with the minimum energy,
using a Lanczos or Davidson algorithm \cite{Lanczos,Davidson,Inexact}. The conventional approach
involves the density matrix renormalization group (DMRG) \cite{white_prl,White,DMRG}. For a
further discussion, we refer to \cite{Inexact}.

The Hadamard product of tensors mentioned in item (b) is important for many
applications, see \cite{Recompression}.  It is well known how to represent a Hadamard product
of two tensors in the TT format as a single tensor in the TT format \cite{Cichocki}.
The problem is that the bond dimension of the product of the two tensors
might be excessively large. For example, if the bond dimensions of the
original tensors are large, such as $B_1 = B_2 = 1000$, then the product
has the bond dimension $B_1B_2 = 10^6$. Each of the middle TT cores (wagons)
of the product tensors would then contain a number of elements on the order
of $10^{12}$, which is too much for customary computers. The beauty of the
algorithm that we propose is that it does not need to work with such huge
arrays but still achieves the desired result, i.e., it finds a rounded TT
representation of the product without the need to evaluate the product tensor
exactly.

The remainder of the paper is organized as follows: Section II compares
TT, CPD, and SOTT models of tensors. Section III presents the algorithm TTSVD-TT0 and TTSVD-TT1.
In Section IV, proposed are extensions of the TTSVD-TT1 algorithm for tensors in different
formats: a SOTT format, a Hadamard product of
two TTs, and the MPO--MPS product. Section V discusses a potential
application of the proposed algorithm for minimizing a quadratic function of
a TT. Section VI provides numerical examples, and Section VII concludes the
paper.

\section{Preliminaries}
\subsection{Tensor Train (TT)}

Consider a general order-$N$ tensor $T$ with elements
$T(i_1, \ldots, i_N)$, where $i_n = 1, \ldots, I_n$ for $n = 1, \ldots, N$.
A tensor train (TT) is a model of a tensor with a moderate number of
parameters arranged in so-called TT-cores or wagons.

The elements of a tensor train are calculated as matrix products
\begin{eqnarray*}
&& T(i_1,\ldots,i_N)=W_1(i_1)W_2(i_2)\ldots W_N(i_N)
\end{eqnarray*}
where $W_n(i_n)$ represents a matrix-valued function of the $n$-th index and has
size $B_{n-1} \times B_n$ for $n = 1, \ldots, N$, with $B_0 = B_N = 1$.
The integers $B_1, \ldots, B_{N-1}$ are called bond dimensions.
Symbolically, we write $T = \{\{W_1, \ldots, W_N\}\}$.

Since $W_n(i_n)$ is a matrix-valued function of the index $i_n$,
it can be understood as an order-3 tensor with dimensions
$B_{n-1} \times I_n \times B_n$.

If all bond dimensions are 1, the tensor becomes the outer product of its wagons,
thus characterizing it as a rank-one tensor.

\subsection{CP Decomposition}

The canonical polyadic decomposition (CPD) decomposes the tensor into a sum
of $R$ rank-one tensors, where $R$ is the rank of the decomposition.
Symbolically, we write
$\tT=[[A_1,\ldots,A_N]]$ where $A_n$, $n=1,\ldots,N$ are factor matrices of the size $I_n\times R$. The notation means that
$$
\tT=\sum_{r=1}^R A_1(:,r) \circ \ldots \circ A_N(:,r)
$$
where $A_n(:, r)$ is the $r$-th column of $A_n$, and $\circ$ denotes the outer (tensor) product.

In this paper, we present an algorithm that allows us to expand any tensor in TT format in a series of
TTs of arbitrarily lower bond dimension. The result is a sum of tensor trains (SOTT). Thus, in the special case of bond dimension 1,
we can convert the tensor from TT format into CPD format.
The other way is also possible. We present an algorithm that enables the conversion of a SOTT into a single TT.
Since CPD is a special case of SOTT, we can convert a tensor in the CPD format into the TT format
without the need to evaluate all particular tensor elements.

\subsection{Error Computation}

\begin{figure}
\begin{center}
\begin{tikzpicture}%
  \useasboundingbox (-9.2pt,12.2pt) rectangle (192.4pt,-41.7pt);
  \foreach \i/\X in {1/0,2/30,3/60,4/90}{
    \node[wagon] (x\i) at (\X pt,  0pt) {$x_{\i}$};
    \node[wagon] (y\i) at (\X pt,-29pt) {$y_{\i}$};
    \draw[bond] (x\i) -- (y\i);
  }
  \node[wagon] (xN) at (180pt,  0pt) {$x_N$};
  \node[wagon] (yN) at (180pt,-29pt) {$y_N$};
  \draw[bond] (xN) -- (yN);
  \foreach \r in {x,y}{
    \draw[bond] (\r1) -- (\r2) -- (\r3) -- (\r4);
    \draw[bond] (\r4) -- ++(24pt,0);
    \draw[bond] (\r N) -- ++(-24pt,0);
  }
  \ttdots{(135pt,0pt)}
  \ttdots{(135pt,-29pt)}
\end{tikzpicture}%
\end{center}
\caption{Scalar product of two tensor trains $x=\{\{x_n\}\}$ and $y=\{\{y_n\}\}$.} 
\label{scaprod}
\end{figure}
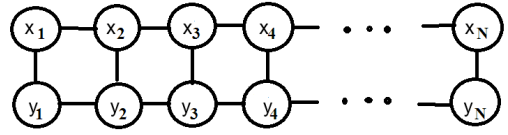

Let $x = \{\{x_n\}\}$ and $y = \{\{y_n\}\}$ represent two tensors in the TT
format having the same order and dimensions.
The scalar product of the tensors, denoted by $\langle x, y\rangle$, can
be computed through tensor contractions \cite{Novikov,Randomized}, as illustrated in Fig. \ref{scaprod}.
In this diagram, the upper row of circles represents the wagons of $x$,
while the lower row represents the wagons of $y$.
The connecting lines indicate the tensor contractions.

Since the structure has no free edges, the result is a scalar.
The contractions are computed either from left to right or from right to left.
In each intermediate step, only matrices of size $B_n \times B_n$ and
reshaped tensor wagons are required.
In MATLAB notation, the algorithm is summarized in Algorithm 1 in the
Supplementary materials or in \cite{code}.\vspace{3mm}

Once we can calculate the scalar product, we can determine the TT norm
$\|x\| = \langle x, x\rangle^{1/2}$ (i.e., the Frobenius norm) or the Frobenius
norm of the difference $x - y$,
$$
\Vert x-y\Vert=\sqrt{\Vert x\Vert^2+\Vert y \Vert^2-2\langle x,y\rangle}~.
$$
In this way, if $y$ is an approximation of $x$ with a lower bond dimension,
we can compute the error of the approximation. The relative error is
$\|x-y\|/\|x\|$.
Similarly, an approximation error for the SOTT or for a tensor in CP format
can also be computed.

\subsection{TT Expansion}

The TT expansion rewrites (or approximates) a tensor train with possibly large
bond dimensions as a SOTT of lower bond dimensions:
\begin{equation}
x=\sum_{m=1}^M u_m + e_M~.
\end{equation}
Here, $x$ is a given tensor train representing a tensor $T$ with maximum bond
dimension $B$. The components $u_m$ $(m=1,\ldots,M)$ are tensor trains of the
same order and dimensions as $x$, but with bond dimensions $B_{\max} < B$.
The error term $e_M$ has a small Frobenius norm, $\|e_M\|=\varepsilon$,
where $\varepsilon$ is small, ideally a machine zero of the computer.
Unlike the other tensors, $e_M$ is not given in the form of a single TT
but is a SOTT:
\begin{equation}
e_M=\{x,-u_1,\ldots,-u_M\}=x-\sum_{m=1}^M u_m~.\label{error}
\end{equation}
Let TTSVD-TT$(x, B_{\max})$ denote the algorithm that approximates a tensor
in TT format with another TT whose bond dimensions are lower than or equal to
$B_{\max}$, and let TTSVD-U$(U, B_{\max})$ denote the algorithm that approximates
a SOTT with a single TT with bond dimension $B_{\max}$. TTSVD-U will be presented
in the next section. Then, the TT expansion can be defined as
\begin{eqnarray}
u_1&=&\mbox{TTSVD-TT}(x,B_{max})\\
u_{m+1}&=&\mbox{TTSVD-U}(\{x,-u_1,\ldots,-u_m\},B_{max})\\ && \rule{0mm}{0mm}\qquad m=1,2,\ldots,M-1~.\nonumber
\end{eqnarray}
The expansion is halted when the fitting error $\varepsilon=\|e_M\|$
becomes sufficiently small, or when the maximum allowed number of terms $M$
is reached. Usually, the Frobenius norms of the components $\|u_m\|$ converge
to zero in the expansion, and the fitting errors $\|e_m\|$ decrease as well.

In the special case $B_{\max}=1$, the method converts the TT format of the tensor
into the CP format. Finally, note that the proposed procedure can be easily
modified for the case where the original tensor $T$ is not given in TT form but
is given as a SOTT.

The tensor expansion is intended to provide correction terms to the TT rounding
operations. Note that TT rounding usually introduces a loss in accuracy; this loss
can be compensated by the expansion.

\section{TT Rounding and TTSVDTT}

The reduction of bond dimensions in tensor train (TT)
models is also known as TT rounding \cite{Ballard, Randomized}. The TT
rounding is performed by truncating singular values between
the tensor ``wagons", as outlined in Algorithm 2.2 of \cite{Randomized}.
The fastest state-of-the-art algorithm is probably the randomized
algorithm using the Khatri--Rao product \cite{Adaptive}. We refer to it as
TTroundingKRP. In this section, we explain the TT rounding
using the TTSVD algorithm for tensors in the TT format,
referred to as TTSVDTT, which was proposed in \cite{Optim}, and
modify it a bit. Our simulations show that the performance
of the original TT rounding and TTSVDTT --meaning the
accuracy in the case of lossy rounding-- is quite similar;
only the latter one is faster in the new implementation.
The algorithm TTroundingKRP is usually faster than TTrounding
and TTSVDTT, but it exhibits a lower precision in the lossy
scenarios.


\subsection{TTSVD Algorithm}

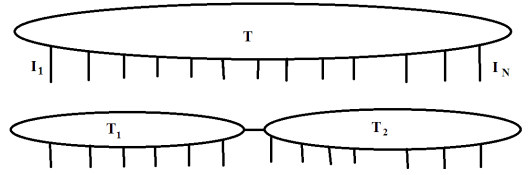
\begin{figure}
\begin{center}
\begin{tikzpicture}%
  \useasboundingbox (0pt,0pt) rectangle (201.6pt,-66.9pt);
  \ttellipse{101.9}{-15}{93.3}{9.9}{$T$}
  \ttlegs{101.9}{-15}{93.3}{9.9}{13}{22.4}{183.3}{-34.5}
  \node at (14pt,-29pt)  {$I_1$};
  \node at (192pt,-29pt) {$I_N$};
  \ttellipse{51.1}{-52}{43.9}{7}{$T_1$}
  \ttellipse{150.7}{-52}{48.4}{7}{$T_2$}
  \draw[bond] (95pt,-52pt) -- (102.3pt,-52pt);
  \ttlegs{51.1}{-52}{43.9}{7}{6}{22.4}{87.3}{-66.4}
  \ttlegs{150.7}{-52}{48.4}{7}{7}{110}{183.3}{-66.4}
\end{tikzpicture}%
\end{center}
\caption{Strangulation of the order-$N$ tensor in two tensors of the smaller size.}
\label{delim}
\end{figure}

The TTSVD algorithm is best explained through the concept
of ``strangulation". Assume that a tensor $T$ of order $N$ and
size $I_1 \times \cdots \times I_N$ is to be rewritten as a
contraction of two smaller tensors, $T_1$ and $T_2$, see Fig. \ref{delim}.
Here, the free edges denote free indices of the structures.

The strangulation is performed through the SVD. First, the
tensor $T$ is reshaped into a matrix $X$ of size
$(I_1 I_2 \cdots I_k)\times(I_{k+1}\cdots I_N)$ and the matrix
is decomposed via the SVD, $X = U S V^{T}$.
Then, tensor $T_1$ is obtained by reshaping $U$ into the size
$(I_1 \times \cdots \times I_k \times B)$, and $T_2$ is obtained
by reshaping $S V^T$ into the size
$(B \times I_{k+1} \times \cdots \times I_N)$, where $B$ is the number
of significant singular values of $X$. Then, $B$ becomes the bond dimension
between $T_1$ and $T_2$.

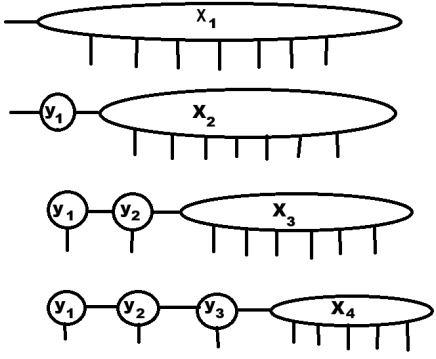
\begin{figure}
\begin{center}
\begin{tikzpicture}%
  \useasboundingbox (0pt,0pt) rectangle (201.6pt,-149.1pt);
  \draw[bond] (17pt,-18pt) -- (29pt,-18pt);
  \ttellipse{97.3}{-18}{68.3}{6.9}{$X_1$}
  \ttlegs{97.3}{-18}{68.3}{6.9}{8}{45}{150}{-34.5}   
  \node[wagon] (a1) at (40pt,-52pt) {$y_1$};
  \draw[bond] (a1) -- ++(-20pt,0);
  \draw[bond] (a1) -- (52.2pt,-52pt);
  \draw[bond] (a1) -- (40pt,-69pt);                  
  \ttellipse{107.7}{-52}{55.5}{8.6}{$X_2$}
  \ttlegs{107.7}{-52}{55.5}{8.6}{7}{65.3}{150}{-69}
  \node[wagon] (b1) at (40pt,-90pt) {$y_1$};
  \node[wagon] (b2) at (66pt,-90pt) {$y_2$};
  \draw[bond] (b1) -- (b2) -- (82.6pt,-90pt);
  \foreach \n in {b1,b2} \draw[bond] (\n) -- (\n |- 0,-107.2pt);
  \ttellipse{125.7}{-90}{43.1}{6.9}{$X_3$}
  \ttlegs{125.7}{-90}{43.1}{6.9}{6}{94.9}{155.8}{-107.2}
  \node[wagon] (c1) at (40pt,-126pt) {$y_1$};
  \node[wagon] (c2) at (66pt,-126pt) {$y_2$};
  \node[wagon] (c3) at (96pt,-126pt) {$y_3$};
  \draw[bond] (c1) -- (c2) -- (c3) -- (116.8pt,-126pt);
  \foreach \n in {c1,c2,c3} \draw[bond] (\n) -- (\n |- 0,-141.7pt);
  \ttellipse{147.1}{-126}{30.3}{7.4}{$X_4$}
  \ttlegs{147.1}{-126}{30.3}{7.4}{5}{125.7}{168.3}{-141.7}
\end{tikzpicture}%
\end{center}
\caption{TT construction through successive $N-1$ strangulations: TTSVD.}
\label{TTSVD0}
\end{figure}

The TTSVD algorithm consists of $N-1$ strangulations with
$k=1$, see Fig. \ref{TTSVD0}. The matrix $X_1$ is obtained by reshaping
the tensor into a matrix of size $I_1 \times (I_2 \cdots I_N)$ and
computing the SVD $X_1 = U S V^{T}$. The first wagon of the TT is obtained
by reshaping $U$ into the shape $1 \times I_1 \times B_1$.
The rest of the tensor is obtained by reshaping the matrix
$S V^{T} = U^{T} X_1$ into another matrix $X_2$ of shape
$(B_1 I_2)\times(I_3 \cdots I_N)$, and so on. There are $N-1$ steps.
The trade-off between the complexity of the model and the accuracy of the
tensor approximation is controlled by the bond dimensions
$B_1,\ldots,B_{N-1}$. Alternatively, it is possible to estimate the
bond dimensions adaptively, based on the numbers of singular values of
the matrices $X_n$ that exceed a tolerance limit.

\subsection{TTSVD-TT0 Algorithm}

Assume that the original TT is $x = \{\{x_n\}\}$ with bond dimensions
$B_n$, $n=0,\ldots,N$. The desired TT, with possibly lower bond dimensions
$B'_n$, $n=0,\ldots,N$, is $y = \{\{y_n\}\}$. The tensors have size
$I_1 \times \cdots \times I_N$.

\begin{figure}
\begin{center}
\begin{tikzpicture}%
  \useasboundingbox (0pt,0pt) rectangle (201.6pt,-44.4pt);
  \node at (7pt,-26pt) {$X_1$};
  \node[wagon] (a) at (47.5pt,-26pt) {$x_1$};
  \node[wagon] (b) at (79pt,-26pt) {$x_2$};
  \node[wagon] (c) at (113pt,-26pt) {$x_3$};
  \node[wagon] (N) at (181pt,-26pt) {$x_N$};
  \ttstub{a}{west}{-9}{$I_1$}
  \bondlab{a}{b}{$B_1$}
  \bondlab{b}{c}{$B_2$}
  \ttstub{c}{east}{13}{$B_3$}
  \ttstub{N}{west}{-11}{$B_{N\ttm 1}$}
  \ttdots{(150pt,-26pt)}
  \legdown{b}{-44}{$I_2$}
  \legdown{c}{-44}{$I_3$}
  \legdown{N}{-44}{$I_N$}
\end{tikzpicture}%

\begin{tikzpicture}%
  \useasboundingbox (0pt,0pt) rectangle (201.6pt,-44.4pt);
  \node at (7pt,-26pt) {$X_2$};
  \node[wagon] (a) at (47.5pt,-26pt) {$z_2$};
  \node[wagon] (b) at (79pt,-26pt) {$x_3$};
  \node[wagon] (c) at (113pt,-26pt) {$x_4$};
  \node[wagon] (N) at (181pt,-26pt) {$x_N$};
  \draw[bond] (a) -- (47.5pt,-8pt) node[idx,above]{$I_2$};
  \ttstub{a}{west}{-9}{$B'_1$}
  \bondlab{a}{b}{$B_2$}
  \bondlab{b}{c}{$B_3$}
  \ttstub{c}{east}{13}{$B_4$}
  \ttstub{N}{west}{-11}{$B_{N\ttm 1}$}
  \ttdots{(150pt,-26pt)}
  \legdown{b}{-44}{$I_3$}
  \legdown{c}{-44}{$I_4$}
  \legdown{N}{-44}{$I_N$}
\end{tikzpicture}%

\begin{tikzpicture}%
  \useasboundingbox (0pt,0pt) rectangle (201.6pt,-44.4pt);
  \node at (7pt,-26pt) {$X_n$};
  \node[wagon] (a) at (47.5pt,-26pt) {$z_n$};
  \node[wwagon] (b) at (79pt,-26pt) {$x_{n\ttp 1}$};
  \node[wwagon] (c) at (113pt,-26pt) {$x_{n\ttp 2}$};
  \node[wagon] (N) at (181pt,-26pt) {$x_N$};
  \draw[bond] (a) -- (47.5pt,-8pt) node[idx,above]{$I_n$};
  \ttstub{a}{west}{-9}{$B'_{n\ttm 1}$}   
  \bondlab{a}{b}{$B_n$}
  \bondlab{b}{c}{$B_{n\ttp 1}$}   
  \ttstub{c}{east}{13}{$B_{n\ttp 2}$}
  \ttstub{N}{west}{-11}{$B_{N\ttm 1}$}
  \ttdots{(150pt,-26pt)}
  \legdown{b}{-44}{$I_{n\ttp 1}$}
  \legdown{c}{-44}{$I_{n\ttp 2}$}
  \legdown{N}{-44}{$I_N$}
\end{tikzpicture}

\begin{tikzpicture}%
  \useasboundingbox (0pt,0pt) rectangle (201.6pt,-44.4pt);
  \node at (12pt,-22pt) {$z_2$};
  \node[wagon] (a) at (71pt,-22pt)  {$y_1$};
  \node[wagon] (b) at (100pt,-22pt) {$x_1$};
  \node[wagon] (c) at (131pt,-22pt) {$x_2$};
  \ttstub{a}{west}{-9}{$B'_1$}
  \bondlab{a}{b}{$I_1$}
  \bondlab{b}{c}{$B_1$}
  \ttstub{c}{east}{13}{$B_2$}
  \legdown{c}{-42}{$I_2$}
\end{tikzpicture}%

\begin{tikzpicture}%
  \useasboundingbox (0pt,0pt) rectangle (201.6pt,-43.8pt);
  \node at (12pt,-22pt) {$z_3$};
  \node[wagon] (a) at (71pt,-22pt)  {$y_2$};
  \node[wagon] (b) at (121pt,-22pt) {$z_2$};
  \node[wagon] (c) at (157pt,-22pt) {$x_3$};
  \ttstub{a}{west}{-9}{$B'_2$}
  \draw[bond] (a.40) to[out=25,in=155,looseness=0.8]   node[idx,above]{$I_2$} (b.140);
  \draw[bond] (a.-40) to[out=-25,in=-155,looseness=0.8] node[idx,below]{$B'_1$} (b.-140);
  \bondlab{b}{c}{$B_2$}
  \ttstub{c}{east}{11}{$B_3$}
  \legdown{c}{-42}{$I_3$}
\end{tikzpicture}%

\begin{tikzpicture}%
  \useasboundingbox (0pt,0pt) rectangle (201.6pt,-43.8pt);
  \node at (12pt,-22pt) {$z_n$};
  \node[wwagon] (a) at (71pt,-22pt)  {$y_{n\ttm 1}$};
  \node[wwagon] (b) at (121pt,-22pt) {$z_{n\ttm 1}$};
  \node[wagon] (c) at (157pt,-22pt) {$x_n$};
  \ttstub{a}{west}{-9}{$B'_{n\ttm 1}$}
  \draw[bond] (a.40) to[out=25,in=155,looseness=0.8]   node[idx,above]{$I_{n\ttm 1}$} (b.140);
  \draw[bond] (a.-40) to[out=-25,in=-155,looseness=0.8] node[idx,below]{$B'_{n\ttm 2}$} (b.-140);
  \bondlab{b}{c}{$B_{n\ttm 1}$}   
  \ttstub{c}{east}{11}{$B_n$}
  \legdown{c}{-42}{$I_n$}
\end{tikzpicture}%





\end{center}
\caption{Matrices $X_1$, $X_2$, $X_n$, and tensors $z_2$, $z_3$, and $z_n$, $n=2,\ldots,N$ in TTSVD applied to a tensor in the TT format.}
\label{ttsvd}
\end{figure}
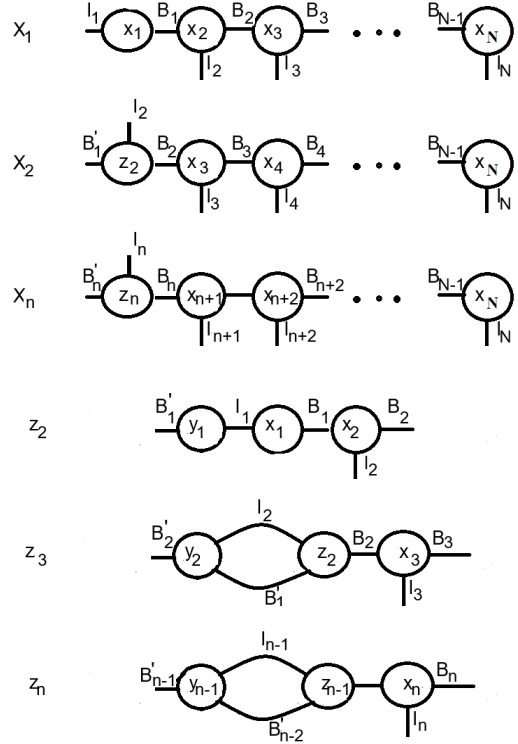

The matrices $X_n$ appearing in the TTSVD can be expressed through the
wagons $x_n$ of the original TT, and auxiliary wagons $z_n$, $n=1,\ldots,N$,
as is shown in Fig. \ref{ttsvd}. The matrix $X_1$ is formed of wagons $x_n$ only.
The first wagon $y_1$ of the new tensor is obtained by
the truncated SVD of $X_1 = U_1 S_1 V_1^{T}$ by reshaping $U_1$.
The matrix $X_2$ is obtained by reshaping the matrix
$X^\prime_2 = S_1 V_1^{T} = U_1^{T} X_1$.
In other words, $X_2$ is obtained by multiplying $y_1$ with $X_1$.
Thus, the matrix $X_2$ has a similar structure to $X_1$: it is formed
of the same wagons as $X_1$ except the first one, which is denoted $z_2$:
\[
X_2 = \mbox{reshape}(\{\{z_2, x_3, \ldots, x_N\}\},B_1^\prime I_2,[\,]).
\]
The tensor (wagon) $z_2$ is obtained as a contraction (product) of $y_1$, $x_1$, and
$x_2$, see Fig. \ref{ttsvd}. Similarly, the tensors $z_n$ and matrices $X_n$
are obtained recursively for $n=2,3,\ldots,N$. Symbolically,
\[
z_n=y_{n-1}\times_{2,1}\times_{3,2}\,z_{n-1}\times_{3,1}x_{n}
\]
where the notation $X \times_{p,q} Y$ denotes the contraction of two tensors $X$ and $Y$ along dimension $p$ of $X$ and dimension $q$ of $Y$; this is defined only if the $p$-th index of $X$ and the $q$-th index of $Y$ have the same range. Similarly, $X \times_{p,q}\times_{r,s} Y$ denotes the contraction of the product along dimensions $p, r$ in $X$ and $q, s$ in $Y$.

The TTSVD-TT0 algorithm does not operate directly on
the matrix $X_n$, because it might have too many elements.
Instead, it operates on the product $X_n X_n^T$ without explicitly
forming $X_n$. The left singular vectors of $X_n$ are derived as
eigenvectors of the matrix $X_n X_n^T$.
This matrix has size $(B_{n-1}^\prime I_n)\times(B_{n-1}^\prime I_n)$, where
$B_{n-1}^\prime$ is the bond dimension between wagons $n-1$ and $n$ in the output tensor, and
$I_n$ is the $n$-th dimension of the tensor, see Fig. \ref{TTSVD-TT}.
Here, $X_n X_n^T$ is computed as a contraction of the wagons
$z_n, x_{n+1}, \ldots, x_N$; each of the wagons appears twice in the contraction graph.
An alternative expression is
\begin{equation}
X_n X_n^T=Z_nM_nZ_n^T \label{prod}
\end{equation}
where $Z_n$ is obtained by re-shaping $z_n$, into a matrix $(IB_{n-1}^\prime)\times B_n$, and
$M_n$ is a $B_n \times B_n$ matrix computed as a contraction of the wagons
$x_{n+1}, \ldots, x_N$. It can be computed recursively, in backward order, for $n = N, N-1, \dots, 1$:
$$
M_n=x_{n+1}\times_{2,1}\times_{3,2} (M_{n+1}\times_{2,3} x_{n+1})
$$
see Fig. 5.
The complexity of the eigendecomposition of $X_n X_n^T$ is $O(I^3 B_{max}^3)$.
This can be prohibitive when the tensor size $I$ is large. Fortunately, an alternative implementation exists.

The eigenvectors of $X_n X_n^T$ can instead be computed as the left singular vectors of a matrix $Y_n$ of size $(I B'_{n-1}) \times B_n$:
$$
Y_n=Z_n\mbox{chol}(M_n)\, ,
$$
where
$S_n=\mbox{chol}(M_n)$ is the Cholesky factor of $M_n$, lower-triangular such that $M_n = S_n S_n^T$.
Note that $X_nX_n^T=Y_nY_n^T$~.
The SVD of $Y_n$ has complexity at most $O(I B_{max} B^2)$, which is linear in $I$. The total complexity of TTSVD-TT0 is therefore
$O(NIB_{max}B^2+NB^3)$. For comparison, the traditional TT-rounding algorithm [2] has complexity $O(NIB^3)$. It follows
that TTSVD-TT0 is faster when $I\gg 1$ and $B_{max}\ll B$.


The algorithm takes two input parameters: a tolerance for rank estimation, and the desired maximum bond dimension $B_{max}$. It is summarized in Table 1, and a MATLAB implementation is provided in the Supplementary Materials.\vspace{3mm}

\noindent {\bf Algorithm 1}: TTSVD-TT0\\
{\bf function} $\{\{y_i\}\}=\mbox{TTSVD-TT0}(\{\{x_i\}\},tol,B_{max})$
\begin{enumerate}
\item $M_N=1$;
\item {\bf for} $n=N-1:-1:1$
\item \rule{0mm}{0mm}\hspace{5mm} $M_n=x_{n+1}\times_{2,1}\times_{3,2} (M_{n+1}\times_{2,3} x_{n+1})$,\quad see Fig.5
\item {\bf end}
\item $z_1=x_1$;
\item {\bf for} $n=1:N-1$
\item  \rule{0mm}{0mm}\hspace{5mm}   $Z_n=\mbox{reshape}(z_n, [\,],B_n)$
\item  \rule{0mm}{0mm}\hspace{5mm}   $[U,\thicksim,\thicksim]=\mbox{svd}(Z_n\mbox{chol}(M_n),tol,B_{max});$
\item  \rule{0mm}{0mm}\hspace{5mm}   $y_n=\mbox{reshape}(U,[B_{n-1}^\prime,I_n,B_n^\prime])$
\item  \rule{0mm}{0mm}\hspace{5mm}   $z_{n+1}=y_n\times_{2,1}\times_{3,2}\,z_n\times_{3,1}x_{n+1}$,\quad see Fig.4
\item {\bf end}
\item $y_N=z_N$
\end{enumerate}

\subsection{New TTSVD-TT Algorithm}

In this paper, we introduce an additional innovation that is more efficient than TTSVD-TT0 when the input bond dimension $B$ is large -- for example, when the input tensor is given as the Hadamard product of two TT-format tensors. It is denoted TTSVD-TT1.

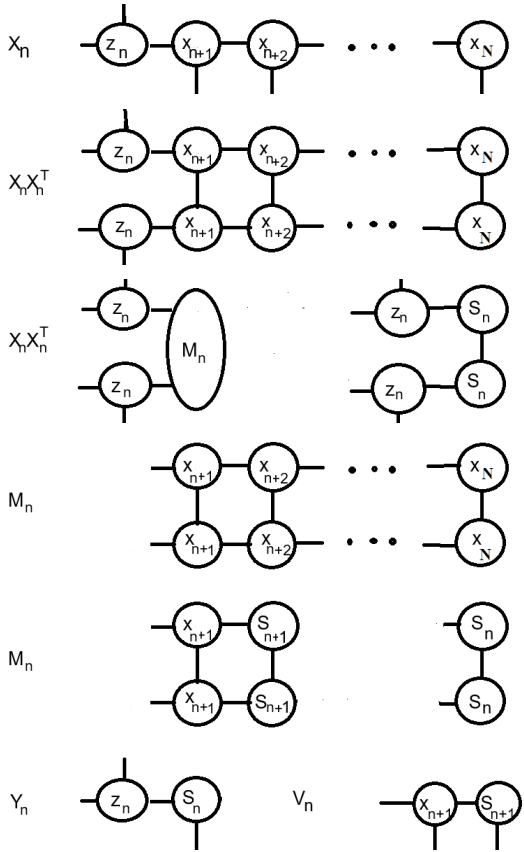
\begin{figure}
\begin{center}
\begin{tikzpicture}%
  \useasboundingbox (-20pt,0pt) rectangle (201.6pt,-33.6pt);
  \node[anchor=west] at (-19pt,-17pt) {$X_n$};
  \node[wagon] (n0) at (47.5pt,-17pt) {$z_n$};
  \node[wwagon] (n1) at (79pt,-17pt) {$x_{n\ttp 1}$};
  \node[wwagon] (n2) at (113pt,-17pt) {$x_{n\ttp 2}$};
  \node[wagon] (n3) at (181pt,-17pt) {$x_N$};
  \draw[bond] (n0) -- (n1);
  \draw[bond] (n1) -- (n2);
  \ttstub{n0}{west}{-9}{}
  \ttstub{n2}{east}{13}{}
  \ttstub{n3}{west}{-11}{}
  \ttdots{(150pt,-17pt)}
  \draw[bond] (n0) -- (n0 |- 0,0pt);
  \draw[bond] (n1) -- (n1 |- 0,-33.6pt);
  \draw[bond] (n2) -- (n2 |- 0,-33.6pt);
  \draw[bond] (n3) -- (n3 |- 0,-33.6pt);
\end{tikzpicture}%
\vspace*{1mm}

\begin{tikzpicture}%
  \useasboundingbox (-20pt,0pt) rectangle (201.6pt,-58.4pt);
  \node[anchor=west] at (-19pt,-30pt) {$X_nX_n^T$};
  \node[wagon] (n0t) at (47.5pt,-16pt) {$z_n$};
  \node[wwagon] (n1t) at (79pt,-16pt) {$x_{n\ttp 1}$};
  \node[wwagon] (n2t) at (113pt,-16pt) {$x_{n\ttp 2}$};
  \node[wagon] (n3t) at (181pt,-16pt) {$x_N$};
  \draw[bond] (n0t) -- (n1t);
  \draw[bond] (n1t) -- (n2t);
  \ttstub{n0t}{west}{-9}{}
  \ttstub{n2t}{east}{13}{}
  \ttstub{n3t}{west}{-11}{}
  \ttdots{(150pt,-16pt)}
  \node[wagon] (n0b) at (47.5pt,-44pt) {$z_n$};
  \node[wwagon] (n1b) at (79pt,-44pt) {$x_{n\ttp 1}$};
  \node[wwagon] (n2b) at (113pt,-44pt) {$x_{n\ttp 2}$};
  \node[wagon] (n3b) at (181pt,-44pt) {$x_N$};
  \draw[bond] (n0b) -- (n1b);
  \draw[bond] (n1b) -- (n2b);
  \ttstub{n0b}{west}{-9}{}
  \ttstub{n2b}{east}{13}{}
  \ttstub{n3b}{west}{-11}{}
  \ttdots{(150pt,-44pt)}
  \draw[bond] (n0t) -- (n0t |- 0,0pt);
  \draw[bond] (n0b) -- (n0b |- 0,-58.4pt);
  \draw[bond] (n1t) -- (n1b);
  \draw[bond] (n2t) -- (n2b);
  \draw[bond] (n3t) -- (n3b);
\end{tikzpicture}%
\vspace{1mm}

\begin{tikzpicture}%
  \useasboundingbox (-20pt,0pt) rectangle (201.6pt,-53.9pt);
  \node[anchor=west] at (-19pt,-27pt) {$X_nX_n^T$};
  \draw[bond] (56.5pt,-14.5pt) -- (71.2pt,-14.5pt);
  \draw[bond] (56.5pt,-39.5pt) -- (71.2pt,-39.5pt);
  \node[tallwagon=40pt, minimum width=20pt] (M) at (79pt,-27pt) {$M_n$};
  \node[wagon] (zt) at (47.5pt,-14.5pt) {$z_n$};
  \node[wagon] (zb) at (47.5pt,-39.5pt) {$z_n$};
  \ttstub{zt}{west}{-9}{}
  \ttstub{zb}{west}{-9}{}
  \draw[bond] (zt) -- (zt |- 0,0pt);   
  \draw[bond] (zb) -- (zb |- 0,-53.9pt);
  \node[wagon] (rt) at (150pt,-14.5pt) {$z_n$};
  \node[wagon] (rb) at (150pt,-39.5pt) {$z_n$};
  \node[wagon] (st) at (181pt,-14.5pt) {$S_n$};
  \node[wagon] (sb) at (181pt,-39.5pt) {$S_n$};
  \ttstub{rt}{west}{-9}{}
  \ttstub{rb}{west}{-9}{}
  \draw[bond] (rt) -- (st);
  \draw[bond] (rb) -- (sb);
  \draw[bond] (st) -- (sb);
  \draw[bond] (rt) -- (rt |- 0,0pt);
  \draw[bond] (rb) -- (rb |- 0,-53.9pt);
\end{tikzpicture}%
\vspace{1mm}

\begin{tikzpicture}%
  \useasboundingbox (-20pt,0pt) rectangle (201.6pt,-53.9pt);
  \node[anchor=west] at (-19pt,-26.5pt) {$M_n$};
  \node[wwagon] (n1t) at (79pt,-12pt) {$x_{n\ttp 1}$};
  \node[wwagon] (n2t) at (113pt,-12pt) {$x_{n\ttp 2}$};
  \node[wagon] (n3t) at (181pt,-12pt) {$x_N$};
  \draw[bond] (n1t) -- (n2t);
  \ttstub{n1t}{west}{-9}{}
  \ttstub{n2t}{east}{13}{}
  \ttstub{n3t}{west}{-11}{}
  \ttdots{(150pt,-12pt)}
  \node[wwagon] (n1b) at (79pt,-41pt) {$x_{n\ttp 1}$};
  \node[wwagon] (n2b) at (113pt,-41pt) {$x_{n\ttp 2}$};
  \node[wagon] (n3b) at (181pt,-41pt) {$x_N$};
  \draw[bond] (n1b) -- (n2b);
  \ttstub{n1b}{west}{-9}{}
  \ttstub{n2b}{east}{13}{}
  \ttstub{n3b}{west}{-11}{}
  \ttdots{(150pt,-41pt)}
  \draw[bond] (n1t) -- (n1b);
  \draw[bond] (n2t) -- (n2b);
  \draw[bond] (n3t) -- (n3b);
\end{tikzpicture}%
\vspace{1mm}

\begin{tikzpicture}%
  \useasboundingbox (-20pt,0pt) rectangle (201.6pt,-53.9pt);
  \node[anchor=west] at (-19pt,-26.5pt) {$M_n$};
  \node[wwagon] (xt) at (79pt,-12pt) {$x_{n\ttp 1}$};
  \node[wwagon] (st) at (113pt,-12pt) {$S_{n\ttp 1}$};
  \node[wagon] (rt) at (181pt,-12pt) {$S_n$};
  \ttstub{xt}{west}{-9}{}
  \ttstub{rt}{west}{-9}{}
  \draw[bond] (xt) -- (st);
  \node[wwagon] (xb) at (79pt,-41pt) {$x_{n\ttp 1}$};
  \node[wwagon] (sb) at (113pt,-41pt) {$S_{n\ttp 1}$};
  \node[wagon] (rb) at (181pt,-41pt) {$S_n$};
  \ttstub{xb}{west}{-9}{}
  \ttstub{rb}{west}{-9}{}
  \draw[bond] (xb) -- (sb);
  \draw[bond] (xt) -- (xb);
  \draw[bond] (st) -- (sb);
  \draw[bond] (rt) -- (rb);
\end{tikzpicture}%

\begin{tikzpicture}%
  \useasboundingbox (-20pt,0pt) rectangle (201.6pt,-35.7pt);
  \node[anchor=west] at (-19pt,-19pt) {$Y_n$};
  \node[wagon] (z) at (47.5pt,-19pt) {$z_n$};
  \node[wagon] (s) at (79pt,-19pt) {$S_n$};
  \ttstub{z}{west}{-9}{}
  \draw[bond] (z) -- (s);
  \draw[bond] (z) -- (z |- 0,0pt);
  \draw[bond] (s) -- (s |- 0,-35.7pt);
  \node at (112pt,-19pt) {$V_n$};
  \node[wwagon] (x) at (150pt,-19pt) {$x_{n\ttp 1}$};
  \node[wwagon] (r) at (182pt,-19pt) {$S_{n\ttp 1}$};
  \ttstub{x}{west}{-9}{}
  \draw[bond] (x) -- (r);
  \draw[bond] (x) -- (x |- 0,-35.7pt);
  \draw[bond] (r) -- (r |- 0,-35.7pt);
\end{tikzpicture}%
\end{center}

\caption{Matrices $X_n$, $X_nX_n^T$, $M_n$, $S_n$, $Y_n$, and $V_n$ in Algorithm 1.}
\label{TTSVD-TT}
\end{figure}

The key novel idea is to compute $S_n$ not as the Cholesky factor of $M_n$, but differently -- without explicitly computing $M_n$ at all. The number of columns of $S_n$ is upper-bounded by $B_{max2}$, an additional parameter of the procedure that should lie between $B_{max}$ and $B$. Specifically, $S_n$ is computed via the truncated SVD of $V_n$, the product of $x_{n+1}$ and $S_{n+1}$, see Fig. \ref{TTSVD-TT},
\begin{equation}
V_n=\mbox{reshape}(x_{n+1}\times_{3,1} S_{n+1},B_n,[\,])~.
\end{equation}
Note that $V_n$ relates to $M_n$ via $M_n = V_n V_n^T$. The new $S_n$ is obtained through a truncated SVD of $V_n$:
\[
[U'_n, S'_n, V'_n] = \mathrm{svd}(V_n, \varepsilon, B_{\max 2}),
\]
where $\varepsilon$ is an error bound on the singular values of $V_n$.
Let $s_1 \ge s_2 \ge \dots $ denote the singular values of $V_n$. Then the number of significant singular values (i.e., the number of columns of $U'_n$) is
\[
B'_n = \min\left(\{\#n;\ s_n \geq \varepsilon s_1\}, B_{\max 2}\right).
\]
The new $S_n$ is defined as $S_n=U_n^\prime S_n^\prime$.

The algorithm consists of two loops:
    Backward loop: computes $S_N = x_N, V_{N-1}, S_{N-1}, \dots, V_1, S_1$.
    Forward loop: computes $y_1, z_2, Y_2, y_2, z_3, Y_3, \dots, y_N$.

Accuracy can be lost at two points: when truncating the singular values of $Y_n$ and of $V_n$. Specifically, the number of significant (non-truncated) values of $Y_n$ is at most $B_{max}$ -- the desired bond-dimension limit. Similarly, the number of significant values of $V_n$ is limited by $B_{max2}$. We can set $B_{max2} = B_{max}$ to get a fast algorithm or $B_{max2} = B$ to get a safer algorithm. If the second option is selected, the algorithm becomes numerically equivalent to TTSVD-TT0. In simulation section we present one example, where the choice $B_{max2} = B_{max}$ is not sufficient to achieve the optimum performance. Anyway, it is advisable to check the rounding accuracy and, if unsatisfactory, to select a larger $B_{max2}$.

The complexity of TTSVD-TT1 is determined by the complexity of the SVD of the matrices $V_n$.
The size of $V_n$ is at most $B_n \times (I_n B_{max2})$. Two cases arise:
Case (1) with $B_n < (I_n B_{max2})$ and complexity $O(B^2 I B_{max2})$, and case (2) with
$B_n \ge (I_n B_{max2})$ and complexity	$O(B I^2 B_{max2}^2)$.
The total complexity of TTSVD-TT1 is therefore $O(N \min\{B^2 I B_{max2}, B I^2 B_{max2}^2\})$.
For fixed $I$ and $B_{max2}$, the complexity is linear in $B$.

A summary of TTSVD-TT1 is presented as pseudocode below. The reshape operator acts on a tensor $X$, transforming it into another tensor with the same number of elements but a specified size -- as in MATLAB. The MATLAB code is provided in the Supplementary Materials or in
\cite{code}.\vspace{3mm}

\noindent {\bf Algorithm 2}: TTSVD-TT1\\
{\bf function} $\{\{y_i\}\}=\mbox{TTSVDTT1}(\{\{x_i\}\},tol,B_{max},B_{max2})$
\begin{enumerate}
\item $S_N=x_N$;
\item {\bf for} $n=N-1:-1:1$
\item \rule{0mm}{0mm}\hspace{5mm} $V_n=\mbox{reshape}(x_{n+1}\times_{3,1} S_{n+1},B_n,[\,])$,\quad see Fig.5
\item  \rule{0mm}{0mm}\hspace{5mm} $[U,S,\thicksim]=\mbox{svd}(V_n,tol,B_{max2});$
 \item \rule{0mm}{0mm}\hspace{5mm} $S_n=US$
\item {\bf end}
\item $z_1=x_1$;
\item {\bf for} $n=1:N-1$
\item  \rule{0mm}{0mm}\hspace{5mm}  $Y_n=\mbox{reshape}(z_n,[\,],B_n)S_n$\quad see Fig.5
\item  \rule{0mm}{0mm}\hspace{5mm}   $[U,\thicksim,\thicksim]=\mbox{svd}(Y_n,tol,B_{max});$
\item  \rule{0mm}{0mm}\hspace{5mm}   $y_n=\mbox{reshape}(U,[B_{n-1},I_n,B_n])$
\item  \rule{0mm}{0mm}\hspace{5mm}   $z_{n+1}=y_n\times_{2,1}\times_{3,2}\,z_n\times_{3,1}x_{n+1}$,\quad see Fig.4
\item {\bf end}
\item $y_N=z_N$
\end{enumerate}

A further modification is possible: Step 4 of the algorithm can be replaced by the operation
$$
[U,S^2,\sim]=\mbox{eig}(W_n,tol,B_{max2}^2)~,
$$
where $W_n=V_nV_n^T$. This approach is subsequently employed in Algorithms TTSVD-U and TTSVD-CP.
It can reduce the memory requirements when $B$ is large, without affecting the asymptotic computational complexity.

\subsection{Reversed-Order TT and Optimality}

A TT model of a tensor can easily be converted into a TT model of the same tensor with the reverse order of modes. This is achieved by reversing the order of the wagons and reversing the order of modes within each wagon -- a computationally inexpensive operation.

This reverse-order operation can be combined with tensor rounding: the tensor is first order-reversed, then rounded to a lower bond dimension, and finally transformed back to its original mode order. This yields a different solution to the TT-rounding task. The two solutions can then be compared, and the one with the lower fitting error (relative to the original tensor) is accepted. Usually, the difference between the two solutions is negligible; however, whether this always holds remains an open theoretical question.

This observation indicates that no form of optimality is currently guaranteed for any TT-rounding procedure. However, there is no reason for concern: the fitting error is easy to check, and if unsatisfactory, one can try TTSVD-TT0, the reversing procedure, or TTSVD-TT1 with a larger $B_{max2}$.

\section{Extensions}

\subsection{TTSVD for tensor in SOTT format}

Let the given SOTT consist of $M$ trains, $U=\{u^{(1)},\ldots,u^{(M)}\}$,
and let $u^{(m)}=\{\{u^{(m)}_1,\ldots,u^{(m)}_N\}\}$ for $m=1,\ldots,M$.
A straightforward implementation of the algorithm would be to convert
the input SOTT into a single TT and then apply TTSVD-TT0.

The SOTT can be represented by a single TT \cite{Cichocki}
$x=\{\{x_1,\ldots,x_N\}\}$ with wagons
\begin{eqnarray*}
x_1&=& [u^{(1)}_1,\ldots,u^{(M)}_1]\\
x_n&=&\left[\begin{array}{ccc} u^{(1)}_n & & 0\\ & \ddots & \\ 0 & & u^{(M)}_n\end{array}\right],
\quad x_N=\left[\begin{array}{c} u^{(1)}_N \\ \vdots \\ u^{(M)}_N\end{array}\right]~.
\end{eqnarray*}
for $n=2,\ldots,N-1$. If $M$ is moderate, this method is likely to be the
most efficient way to implement TTSVD-U. We refer to the algorithm as
TTSVD-U0.

A challenge arises when $M$ is large, for example when $M=1000$.
Since $x_n$ has $O(M^2)$ elements, the memory requirements and the
algorithm's complexity become quadratic functions of $M$.
Therefore, we propose a variant of the TTSVD-TT1 algorithm tailored to this
case that exploits the fact that the wagons are block-diagonal.
Instead of storing the full wagon $x_n$, we store only its diagonal blocks,
and we also compute products involving $x_n$ when building the matrices
$V_n$ and $Y_n$, which is computationally light because we only multiply
with the diagonal blocks. The resulting algorithm is called TTSVD-U and is given in the
Supplementary Materials and on the Internet.
The complexity of TTSVD-U can be deduced from the complexity of TTSVD-TT1.
It is asymptotically proportional to $M$, when the other parameters are fixed.

If the accuracy is not satisfactory, we found it useful to use aggregation
rather than TTSVD-U directly. This means that the TTs are divided into groups,
the TTs in each group are summed separately, and then the partial sums are
added together. We will not discuss this issue further here, as it is
problem-specific.


\subsection{TTSVD for Hadamard product of two TTs}

Assume that the input tensor $x=\{\{x_n\}\}$ is obtained as a Hadamard
(elementwise) product of tensors $f=\{\{f_n\}\}$ and $g=\{\{g_n\}\}$ of the same
dimensions (not necessarily with the same bond dimensions). It is well known \cite{Cichocki}
that the wagons $x_n$ of $x$ have slices
$$
x_n(:,i,:)=f_n(:,i,:)\otimes g_n(:,i,:),\qquad i=1,\ldots,I_n
$$
In other words, the slices of $x_n$ are obtained as the Kronecker product of the
corresponding slices of $f_n$ and $g_n$. If $f$ and $g$ have bond dimensions
$B_f$ and $B_g$, respectively, then the bond dimension of $x$ is $B_fB_g$.
If both $B_f$ and $B_g$ are large, say $1000$, even storing the wagons $x_n$
may not be practical. Instead, we propose a variant of the TTSVD-TT1 algorithm
called TTSVD-HP, in which $x_n$ need not be stored in its full form.
We only need to compute products of $x_n$ with other matrices.
When computing $S_n$, $Y_n$, and $V_n$ in TTSVD-TT1, we can avoid manipulating
$x_n$ directly. Instead of multiplying by $x_n$, we multiply by the wagons of the
original tensors, $f_n$ and $g_n$.

In the first (backward) loop, we compute the matrices $S_n$ recursively as in
TTSVD-TT1. The iteration begins with
\[
S_N=x_N=f_N(:,:,1)\odot g_N(:,:,1),
\]
where $\odot$ denotes the Khatri--Rao product.
For each step $n=N-1,\ldots,1$, $S_n$ is obtained by a truncated SVD of a
matrix $V_n$, given as the product $x_{n+1}S_{n+1}$.
Assume that $S_{n+1}$ has size $(B_fB_g)\times B_{\max}$ and that
$x_{n+1}$ has size $(B_fB_g)\times I_{n+1}\times(B_fB_g)$.

For each $i=1,\ldots,I_n$ we need to compute the product
\begin{eqnarray*}
x_{n+1}(:,i,:)S_{n+1}=[f_{n+1}(:,i,:)\otimes g_{n+1}(:,i,:)]S_{n+1}\\=[I\otimes g_{n+1}(:,i,:)][f_{n+1}(:,i,:)\otimes I]S_{n+1}
\end{eqnarray*}
where $I$ is the identity matrix. The multiplication of $S_{n+1}$ with a
Kronecker product of two matrices can be done in a way inspired by the well-known matrix
multiplication identity
$$
\mbox{vec} (ABC)= (C^T\otimes A) \mbox{vec}{B}
$$
where $A,B,C$ are matrices of compatible dimensions and $\mathrm{vec}$ denotes
the vectorization operator. This identity shows how multiplication with the
``large" matrix $(C^T\otimes A)$ can be performed through multiplication with
``small" matrices $A$ and $C$.

We apply this idea to compute the product
$[f_n(:,i,:)\otimes I]S_n$ by multiplying $f_n(:,i,:)$ (of size $B_f\times B_f$)
with $S_n$ reshaped to the format $B_f\times(B_gB_{\max})$, and then reshaping back.
Similarly, multiplication with the other term $[I\otimes g_n(:,i,:)]$ can be obtained
by multiplying $g_n(:,i,:)$ (of size $B_g\times B_g$) with a matrix of size
$B_g\times(B_fB_{\max})$. In this way, the tensor $V_n$ is computed.
In matrix form, it has size $(B_fB_g)\times(I_{n+1}B_{\max})$.
This matrix is compressed via the SVD into the desired matrix $S_n$.
Note that $V_nV_n^T \approx S_nS_n^T$.

Using the same idea, the tensor $z_n$ is computed; it is the product of $Y_{n-1}$,
$z_{n-1}$, and $x_n$, see Fig. \ref{ttsvd}. The tensor $z_n$ has approximately the same
size, $B_{\max}\times I_n\times(B_fB_g)$.

The resulting algorithm is presented in the Supplementary Materials and on the
Internet. In our MATLAB code, we use the function ``pagemtimes,'' which performs
matrix multiplication of arrays of matrices in parallel with respect to index $i$.

Note that the largest intermediate arrays in the algorithm, $V_n$ and $z_n$,
have size $B_f\times B_g\times B_{\max}\times I_n$.
They are smaller than the size of the wagons of the product tensor,
$(B_fB_g)\times I_n\times(B_fB_g)$, if $B_{\max 2}<B_fB_g$. We do  not even need to compute these wagons.
Complexity of our algorithm TTSVD-HP is linear in $B_fB_g$. The complexity follows the complexity of TTSVD-TT1.
$\min\{O(NB_fB_gI^2B_{\max 2}^2),O(NB_f^2B_g^2IB_{\max 2})\}$~.

\subsection{TTSVD for a matrix--vector product of two TTs}

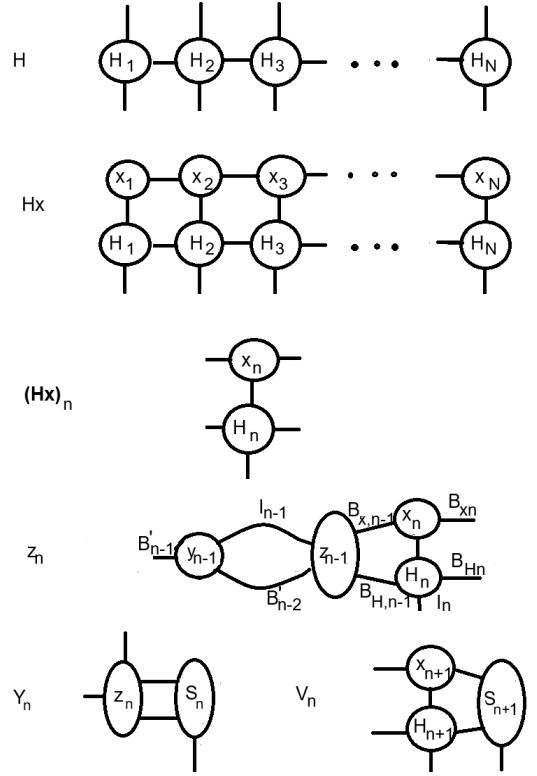
\begin{figure}
\begin{center}
\begin{tikzpicture}%
  \useasboundingbox (0pt,0pt) rectangle (201.6pt,-35.7pt);
  \node at (7pt,-18pt) {$H$};
  \foreach \n/\x/\t in {h1/47.5/H_1, h2/79/H_2, h3/113/H_3, hN/181/H_N}{
    \node[wagon] (\n) at (\x pt,-18pt) {$\t$};
    \draw[bond] (\n) -- (\n |- 0,0pt);
    \draw[bond] (\n) -- (\n |- 0,-35.7pt);
  }
  \draw[bond] (h1) -- (h2) -- (h3);
  \ttstub{h3}{east}{13}{}
  \ttstub{hN}{west}{-11}{}
  \ttdots{(150pt,-18pt)}
\end{tikzpicture}%
\vspace{3mm}

\begin{tikzpicture}%
  \useasboundingbox (0pt,0pt) rectangle (201.6pt,-56pt);
  \node at (9pt,-25pt) {$Hx$};
  \foreach \k/\x/\s in {1/47.5/1, 2/79/2, 3/113/3, N/181/N}{
    \node[wagon] (x\k) at (\x pt,-12pt) {$x_\s$};
    \node[wagon] (h\k) at (\x pt,-38pt) {$H_\s$};
    \draw[bond] (x\k) -- (h\k);
    \draw[bond] (h\k) -- (h\k |- 0,-56pt);
  }
  \foreach \r/\y in {x/-12, h/-38}{
    \draw[bond] (\r1) -- (\r2) -- (\r3);
    \ttstub{\r3}{east}{13}{}
    \ttstub{\r N}{west}{-11}{}
    \ttdots{(150pt,\y pt)}
  }
\end{tikzpicture}%

\begin{tikzpicture}%
  \useasboundingbox (0pt,0pt) rectangle (201.6pt,-64.1pt);
  \node at (17pt,-34pt) {$(Hx)_n$};
  \node[wagon] (x) at (93.5pt,-21pt) {$x_n$};
  \node[wagon] (h) at (93.5pt,-47pt) {$H_n$};
  \draw[bond] (x) -- (h);
  \ttstub{x}{west}{-11}{}
  \ttstub{x}{east}{11}{}
  \ttstub{h}{west}{-11}{}
  \ttstub{h}{east}{11}{}
  \draw[bond] (h) -- (h |- 0,-64.1pt);
\end{tikzpicture}%
\vspace{2mm}

\begin{tikzpicture}%
  \useasboundingbox (0pt,0pt) rectangle (201.6pt,-55.8pt);
  \node at (12pt,-25pt) {$z_n$};
  \node[wwagon] (y) at (64pt,-25pt) {$y_{n\ttm 1}$};
  \node[tallwagon=32pt] (z) at (108pt,-25pt) {$z_{n\ttm 1}$};
  \node[wagon] (x) at (160pt,-12pt) {$x_n$};
  \node[wagon] (h) at (160pt,-38pt) {$H_n$};
  \ttstub{y}{west}{-9}{$B'_{n\ttm 1}$}
  \draw[bond] (y.40) to[out=25,in=155,looseness=0.8]   node[idx,above]{$I_{n\ttm 1}$} (z.150);
  \draw[bond] (y.-40) to[out=-25,in=-155,looseness=0.8] node[idx,below]{$B'_{n\ttm 2}$} (z.-150);
  \draw[bond] (z) -- node[idx,above=1pt,pos=0.5]{$B_{x,n\ttm 1}$} (x);
  \draw[bond] (z) -- node[idx,below=1pt,pos=0.5]{$B_{H,n\ttm 1}$} (h);
  \draw[bond] (x) -- (h);
  \draw[bond] (x.east) -- ++(7pt,0) node[idx,right]{$B_{xn}$};
  \draw[bond] (h.east) -- ++(7pt,0) node[idx,right]{$B_{Hn}$};
  \legdown{h}{-55.8}{$I_n$}
\end{tikzpicture}%

\begin{tikzpicture}%
  \useasboundingbox (0pt,0pt) rectangle (201.6pt,-54.9pt);
  \node at (7pt,-28pt) {$Y_n$};
  \draw[bond] (47.5pt,-22pt) -- (79pt,-22pt);
  \draw[bond] (47.5pt,-34pt) -- (79pt,-34pt);
  \draw[bond] (47.5pt,-28pt) -- (47.5pt,0pt);
  \draw[bond] (79pt,-28pt) -- (79pt,-54.9pt);
  \node[tallwagon=32pt] (z) at (47.5pt,-28pt) {$z_n$};
  \node[tallwagon=32pt] (s) at (79pt,-28pt) {$S_n$};
  \ttstub{z}{west}{-9}{}
  \node at (112pt,-28pt) {$V_n$};
  \draw[bond] (147pt,-14pt) -- (182pt,-28pt) -- (147pt,-42pt);
  \draw[bond] (182pt,-28pt) -- (182pt,-54.9pt);
  \node[tallwagon=34pt, minimum width=20pt] (s1) at (182pt,-28pt) {$S_{n\ttp 1}$};
  \node[wwagon] (x) at (147pt,-14pt) {$x_{n\ttp 1}$};
  \node[wwagon] (h) at (147pt,-42pt) {$H_{n\ttp 1}$};
  \draw[bond] (x) -- (h);
  \ttstub{x}{west}{-9}{}
  \ttstub{h}{west}{-9}{}
  \draw[bond] (h) -- (h |- 0,-54.9pt);
\end{tikzpicture}%
\end{center}

\caption{Matrix $H$ in a TT format and product $Hx$ in the TT format, one wagon of the product tensor, and tensors $z_n$, $Y_n$ and $V_n$ needed in the TTSVD-HX algorithm.}
\label{hami}
\end{figure}

Assume that we are given a matrix $H$ and a tensor $x$ in the TT format,
see Fig. \ref{hami}. We wish to apply the TTSVD-TT1 algorithm to the matrix--tensor product $Hx$.

If the bond dimensions of $H$ and $x$ are $B_H$ and $B_x$, respectively,
then the product $y=Hx$ can be written as a single TT with product bond
dimension $B_HB_x$. The wagons of the product tensor have five free indices,
i.e., five dimensions $B_H\times B_x\times I_n\times B_x\times B_H$,
which is equivalent to the order-3 wagon of size $(B_HB_x)\times I_n\times(B_xB_H)$.

As in the previous section, we will not evaluate the wagons $(Hx)_n$ of the
product tensor. Instead, we compute the
auxiliary matrices $V_n$ and $Y_n$ in the TTSVD-TT1 algorithm without forming these wagons in their full
form. The auxiliary tensors $z_n$, $V_n$, and $Y_n$ are depicted in Fig. \ref{hami}.
Note that the largest intermediate arrays in the algorithm $z_n$, $Y_n$ and $V_n$ have sizes
$B_{\max}\times B_H\times B_x\times I_n$, $B_{\max}\times I_n\times B_{\max}$, and
$B_{\max}\times B_H\times B_x\times I_n$, respectively.
They are smaller than the size of the wagons of the product tensor,
$(B_HB_x)\times I_n\times(B_HB_x)$. In the proposed algorithm TTSVD-HX, the complexity is linear in $B_HB_x$.

Again, the algorithm is presented in the Supplementary Materials and
on the Internet.

\section{Minimizing a quadratic form of a TT}

In quantum chemistry and in other applications, there is a need to minimize a quadratic
function of a tensor that is considered to be in the TT format,
$$
\min \langle x,Hx\rangle
$$
subject to $\|x\|=1$ where $H$ is a symmetric linear operator, often referred to as Hamiltonian.
In the context of quantum chemistry, consider an $N$- particle system described by single-electron integrals $T_{ij}$ and two-electron integrals $V_{ijk\ell}$, $i,j,k,\ell=1,\ldots,N$. Let us introduce three $4\times 4$ matrices
\begin{eqnarray*}
\bc_1=\left(\begin{array}{cccc} 0 & 0 & 1 & 0 \\ 0 & 0 & 0 & 1\\ 0 & 0 & 0 & 0\\ 0 & 0 & 0 & 0
\end{array}\right),\qquad \bc_2=\left(\begin{array}{cccc} 0 & 1 & 0 & 0 \\ 0 & 0 & 0 & 0\\ 0 & 0 & 0 & -1\\ 0 & 0 & 0 & 0\end{array}\right),
\end{eqnarray*}
and $\bz=\mbox{diag} (1,-1,-1,1)$. Next, let us define operators
\begin{eqnarray*}
\ba_{i,s}=\bz\otimes \ldots \otimes\bz\otimes\bc_s\otimes\bI\otimes\ldots\otimes\bI
\end{eqnarray*}
where $\bc_s$ stands at position $i$, $i=1,\ldots, N$, $s=1,2$, and $\bI$ is the $4\times 4$ identity matrix.

The Hamiltonian is defined as \cite{DMRG}
\begin{eqnarray}
H&=&\sum_{i,j=1}^N\sum_{s=1}^2 T_{ij}\ba_{i,s}^\dagger\ba_{j,s} \label{pet} \\&&+\sum_{i,j,k,\ell=1}^N\sum_{s_1,s_2=1}^2 V_{ijk\ell}\ba_{i,s_1}^\dagger\ba_{j,s_2}\ba_{k,s_2}^\dagger\ba_{\ell,s_1}~.\nonumber
\end{eqnarray}
It is a sum of $M=2N^2+4N^4$ terms, each of which represents a rank-one tensor of the size $16\times 16\times\ldots\times 16$ ($16^N$).
If $x$ represents a wave function of the size $4\times 4\times\ldots\times 4$ ($4^N$), then $Hx$ is a tensor of the same shape as $x$.

We begin by converting the Hamiltonian $H$ to TT format using the proposed TTSVD-CP and TTSVD-U algorithms (with aggregation).
Once we have a recipe for computing $y=Hx$, it is possible to minimize
$\langle x,Hx\rangle$ by means of the Lanczos or Davidson algorithm \cite{Lanczos,Davidson,Inexact}. 
In this paper, we only focus on the first part, the MPO--MPS product \cite{Successive}.

\section{Experiments}

\subsection{Performance of TTSVD-TT1}

\begin{figure}
\begin{center}
\includegraphics[width=0.9\linewidth]{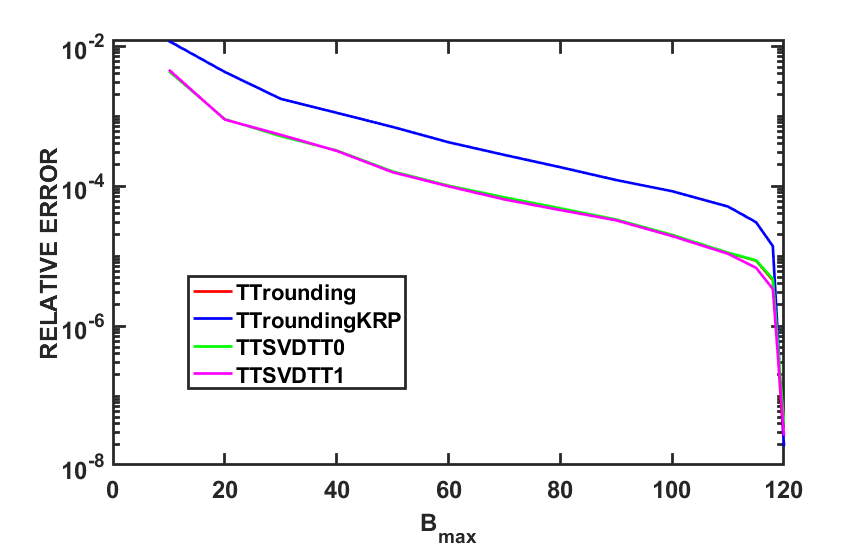}
\includegraphics[width=0.9\linewidth]{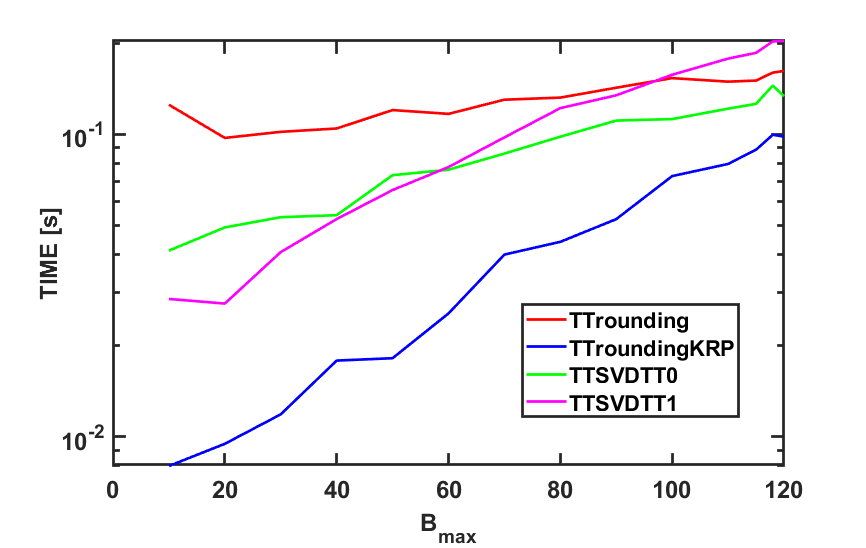}
\includegraphics[width=0.9\linewidth]{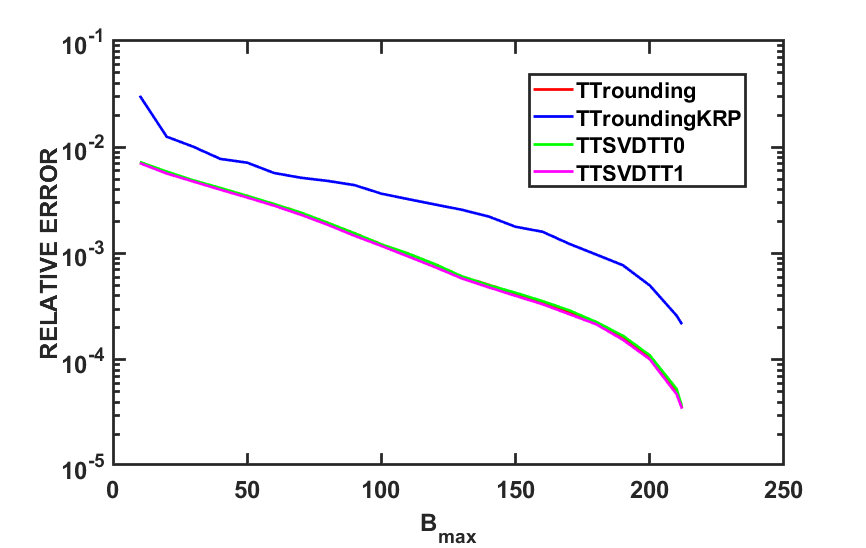}
\includegraphics[width=0.9\linewidth]{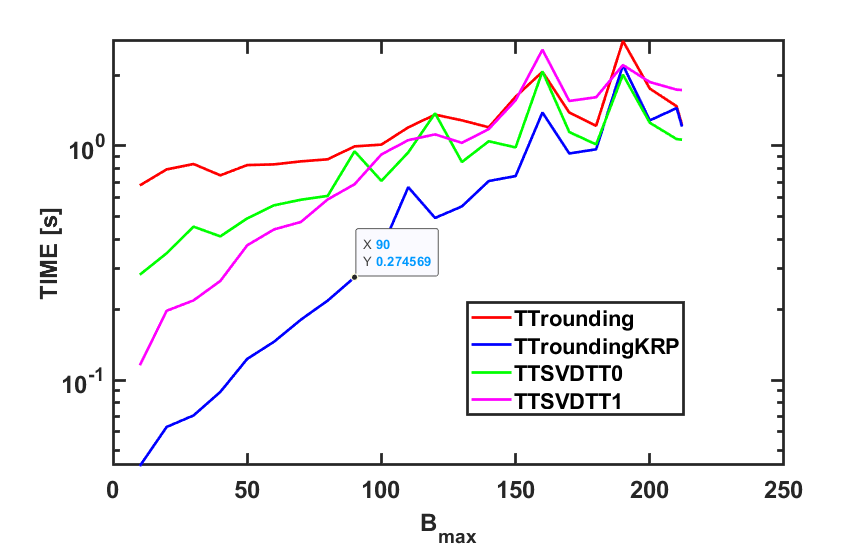}
\end{center}
\caption{Relative error of rounding of chemical tensors (Hamiltonians) of the order 18 and 56, respectively, and computational times.
The fitting errors of TTrounding and TTSVD-TT coincide.}
\label{grafyTT}
\end{figure}

We studied the performance of the TTSVD-TT0, TTSVD-TT1, traditional TT rounding and TTroundingKRP \cite{Adaptive}.
We present two examples: TT rounding of a tensor of the size $16\times\ldots\times 16$
of the order $N=18$ and $N=56$, respectively. The initial bond dimension of these tensors
is $B=120$ and $B=217$, respectively. The origin of these tensors is explained in Sections V and VI.E.
We perform TT rounding of these tensors to a varying (lower) bond dimension, measure the relative fitting errors
$\|T-\hat{T}\|/\|T\|$ and computational time. The speed was measured on a Dell OptiPlex 3090 computer.
The results are shown in Fig. \ref{grafyTT}.

We can see that the fitting errors of TTSVD-TT0 and TTSVD-TT1 closely match that of traditional TT rounding, whereas randomized TT rounding produces an error roughly three times larger. The randomized TT rounding is the fastest method. TTSVD-TT0 and TTSVD-TT1 have nearly the same complexity; the latter method is slightly faster when the target bond dimension $B_{max}$ is small.

This TT rounding approach is motivated by the energy minimization (eigenvalue) problem, which becomes computationally expensive for large bond dimensions. Our strategy is to solve the eigenproblem on a low-bond-dimension approximation first, then progressively increase the bond dimension and update the eigenvectors until the desired "chemical" accuracy is reached.

\subsection{Performance of TTSVD-U}

We studied the performance of the TTSVD-U algorithm in
comparison to TTSVD-U0 and TT rounding. Performance gains
were obtained, namely in the special case in which all
components of the SOTT have bond dimension 1, so that it is a CP model.
Therefore, we designed a specialized algorithm for this
case, denoted TTSVD-CP. In contrast, when the components
did not all have bond dimension 1, TTSVD-U0 using the novel
TTSVD-TT algorithm performed best.

\begin{figure}
\begin{center}
\includegraphics[width=0.9\linewidth]{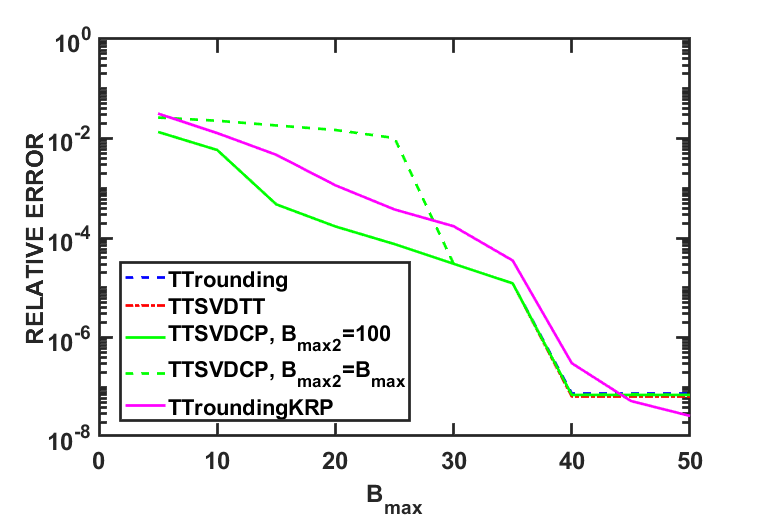}\\
\includegraphics[width=0.9\linewidth]{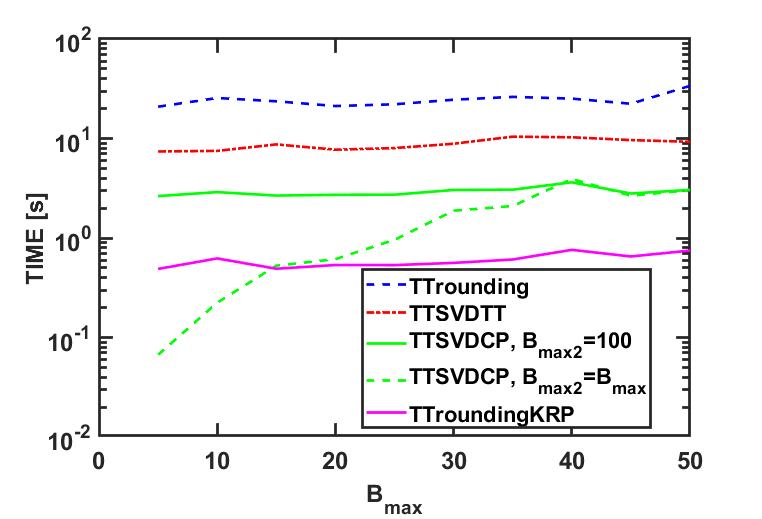}
\end{center}
\caption{Relative error of rounding of a tensor with CP rank 1000 into TT format, and computational time.}
\label{grafyCP}
\end{figure}

For testing the TTSVD-CP algorithm, we considered the
Hamiltonian of [18]annulene, which is introduced in
subsection VI.E. We used the first 1000 most significant rank--1
components of this tensor. The tensor (as well as its
components) has the size $16 \times \cdots \times 16$ ($16^{18}$).
The sum of these 1000 components has the bond dimension
$B = 1000$. We then compressed the tensor using the
algorithms TT rounding, TTSVD-TT1, TTSVD-CP, and TTroundingKRP \cite{Adaptive}
for various bond dimensions, see Fig. \ref{grafyCP}. We can see that the
fastest algorithm -- TTroundingKRP -- has a larger fitting error
in the lossy scenario, $B_{\max} < 40$.
The traditional TT rounding algorithm is the slowest one, and it
has the same fitting error as TTSVD-TT1 and TTSVD-CP. However,
TTSVD-CP achieves the optimum performance only if the parameter
$B_{\max 2}$ is larger than $B_{\max}$. With $B_{\max 2} = 100$,
TTSVD-CP is the second-fastest algorithm among the most accurate algorithms.

\subsection{Performance of TTSVD-HP}

In the first experiment, we consider a Hadamard product
of two identical tensors $F = G$ of order 10 and size
$50 \times 50 \times \cdots \times 50$. The tensor is built from random
wagons with bond dimension $B = 20$. The Hadamard product has,
in theory, bond dimension 400. Practically, it can be compressed
to bond dimension 220 with a negligible fitting error.
We round the tensor to bond dimensions between 20 and 220 and
measure the fitting error and computation error for four algorithms:
TTrounding, TTroundingKRP, TTSVD-TT1, and TTSVD-HP.

\begin{figure}
\begin{center}
\includegraphics[width=0.9\linewidth]{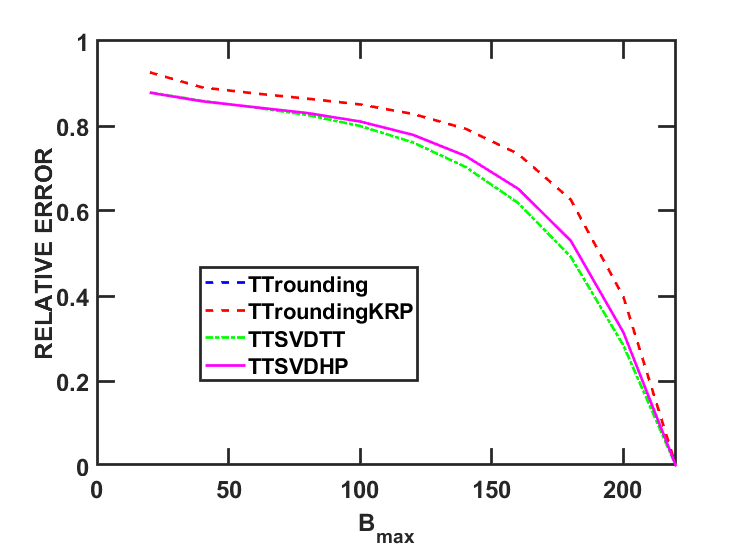}\\
\includegraphics[width=0.9\linewidth]{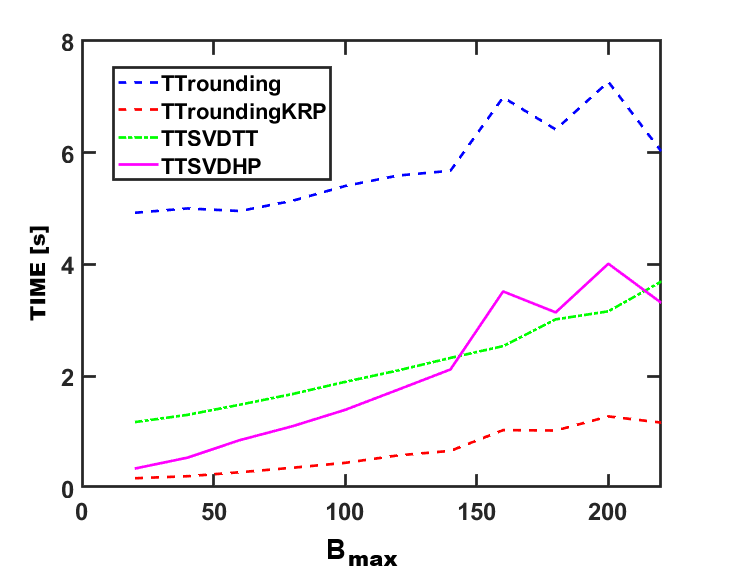}
\end{center}
\caption{Relative error of rounding of an Hadamard product of two TT tensors of the size $50\times \cdots\times 50$ ($50^{10}$) and bond dimension 20, and computational time.}
\label{grafyHP}
\end{figure}

Again, the fastest algorithm, TTroundingKRP, has a larger fitting error
than the remaining three algorithms. Among the other three algorithms,
which achieve nearly the same accuracy, TTSVD-HP and TTSVD-TT are
almost equally fast (both faster than TTrounding), but TTSVD-HP can
handle higher dimensions than the other algorithms because it does not
need to represent the wagons of the product tensor in their full form.

\begin{figure}
\begin{center}
\includegraphics[width=0.9\linewidth]{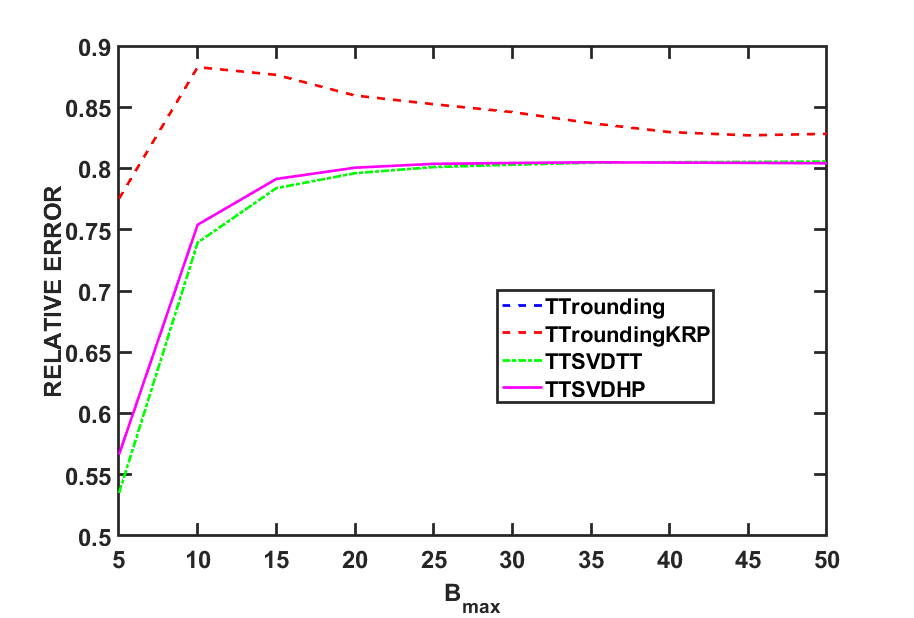}\\
\includegraphics[width=0.9\linewidth]{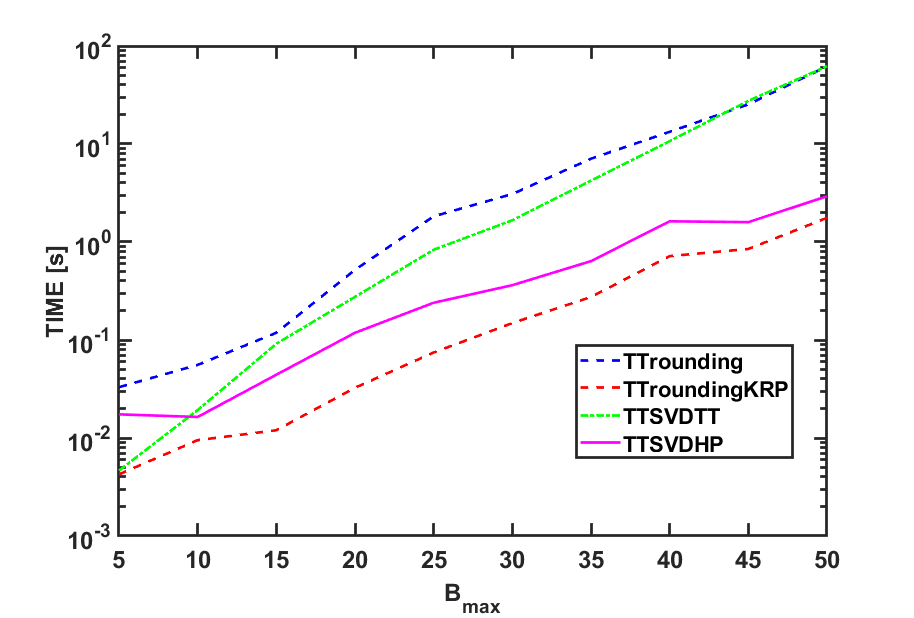}
\end{center}
\caption{Relative error of rounding of an Hadamard product of two TT tensors of the size $4\times \cdots\times 4$ ($4^{18}$) and varying bond dimension, and computational time.}
\label{grafyHP2}
\end{figure}

In the second experiment, we considered a tensor of larger order
and a smaller physical dimension. We generated one TT, $F_0$, of size
$4 \times \cdots \times 4$ ($4^{18}$) with random wagons of bond dimension 100.
The tensor $F$ was obtained by compressing $F_0$ to the bond dimension
$B_{\max}$ between 5 and 50. We then computed the elementwise product
$F * F$ with bond dimension $B_{\max}^2$. This product tensor is
compressed to bond dimension $2B_{\max}$ using the four algorithms again:
TTrounding, TTroundingKRP, TTSVD-TT1, and TTSVD-HP. The first four methods use
the intermediate tensor with bond dimension $B_{max}^2$, the last method does not.
Results are shown in Fig. \ref{grafyHP2}.

Again, the fastest algorithm, TTroundingKRP, has a larger fitting error
than the remaining three algorithms. TTSVD-HP has nearly the same
performance as TTSVD-TT and TTrounding on the product tensor, but it is
much faster, especially for large $B_{\max}$. Moreover, it allows to handle
tensors with larger bond dimension, which the other methods cannot.

\subsection{Performance of TTSVD-HX}

In this experiment, we consider a product of the type $y = Hx$.
The operator $H$ is obtained by compressing the Hamiltonian of
[18]annulene using TTSVD-CP (as described in the next subsection),
and compressing it to bond dimension $B_{\max}$ varying from 5 to 50.
Similarly, $x$ is obtained by compressing the corresponding wave function
to the same bond dimension.

The ground-truth product $y = Hx$ has bond dimension $B_{\max}^2$.
This bond dimension is reduced back to $B_{\max}$ using three algorithms:
TTrounding, TTroundingKRP, TTSVD-TT1, TTSVD-HX, and SRC \cite{Successive}. The results
are presented in Fig. \ref{grafyHX}. We can see that (1) TTrounding, TTSVD-TT1, and
TTSVD-HX have nearly the same performance (the curves overlap), (2) TTroundingKRP
and SRC have a larger fitting error than the other two methods, and
(3) the computational complexity of TTSVD-HX, SRC, and TTroundingKRP grows
at a much lower rate with increasing $B_{\max}$ than that of the other
algorithms.

\begin{figure}
\begin{center}
\includegraphics[width=0.9\linewidth]{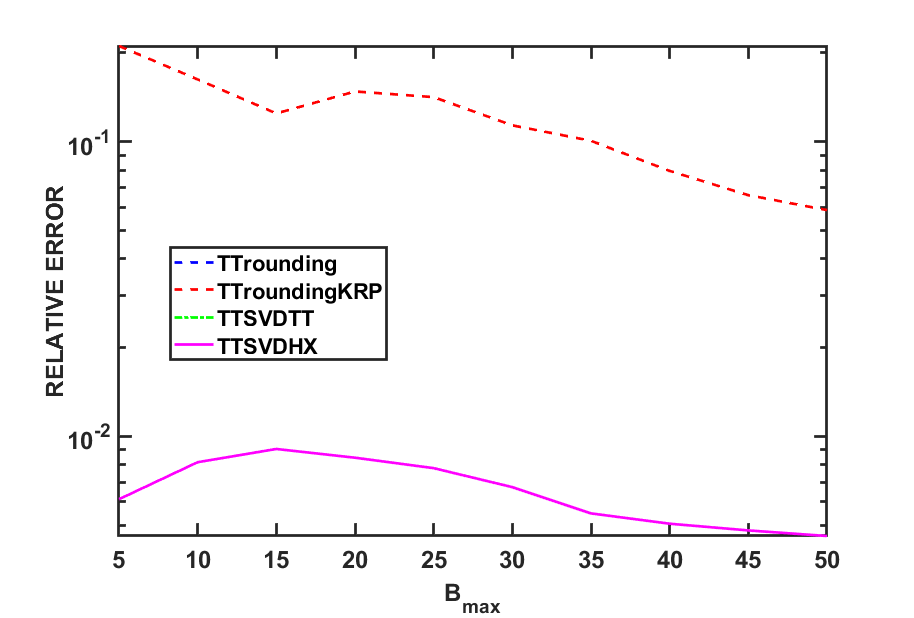}\\
\includegraphics[width=0.9\linewidth]{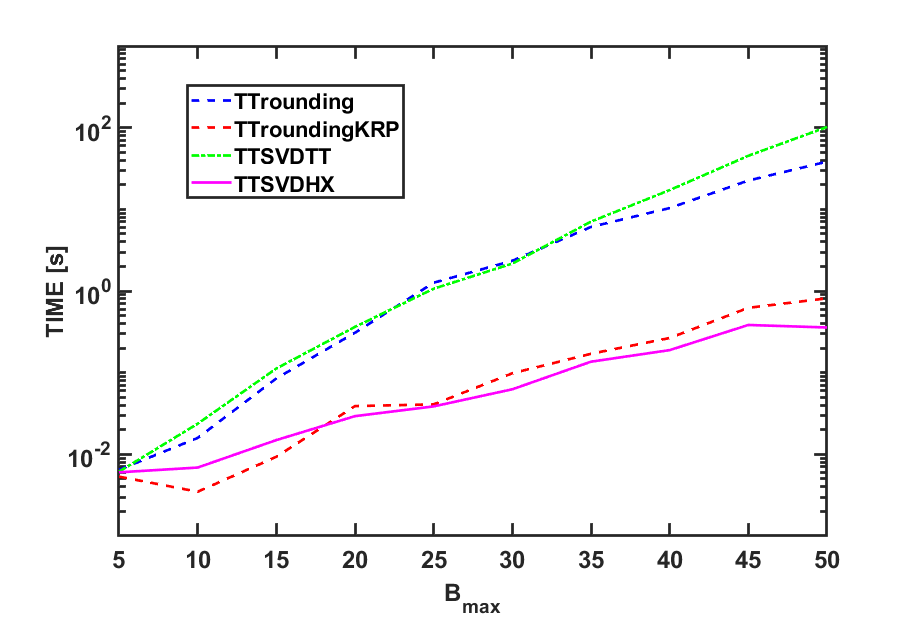}
\end{center}
\caption{Relative error of rounding of product $y=Hx$, and computational time versus the bond dimension of $H$ and $x$.
In the upper diagram, the curves for the relative errors of TTrounding, TTSVD-TT1 and TTSVD-HX overlap.}
\label{grafyHX}
\end{figure}

Finally, we compare the performance of the TTSVD-HX algorithm with SRC in
an experiment proposed in \cite{Successive}, Section 4.4.2. Here, the tensor $X$ (MPS)
has the size $2 \times \cdots \times 2$ ($2^{100}$), the Hamiltonian (MPO)
and the MPS have bond dimension 50, and their wagons contain uniformly random
entries $[\alpha$, 1] with $\alpha = -0.5$.

\begin{figure}
\begin{center}
\includegraphics[width=0.8\linewidth]{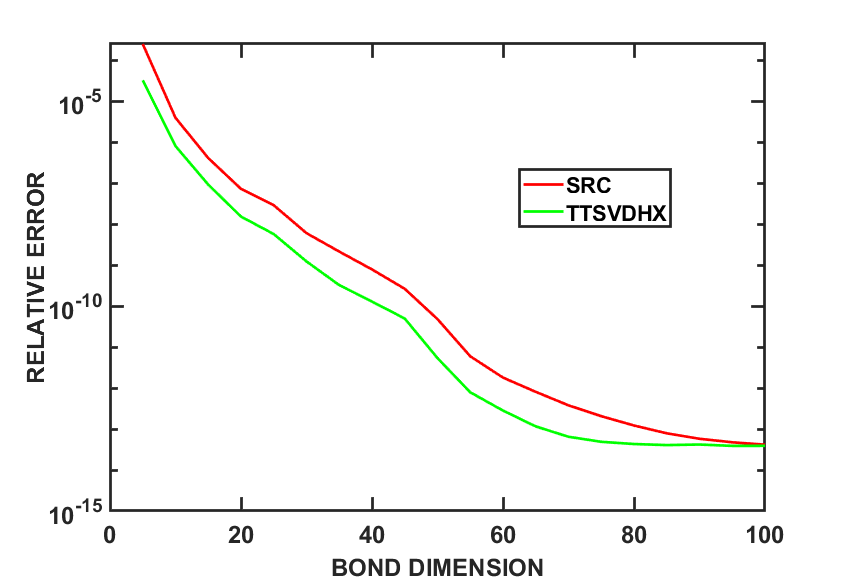}\\
\includegraphics[width=0.8\linewidth]{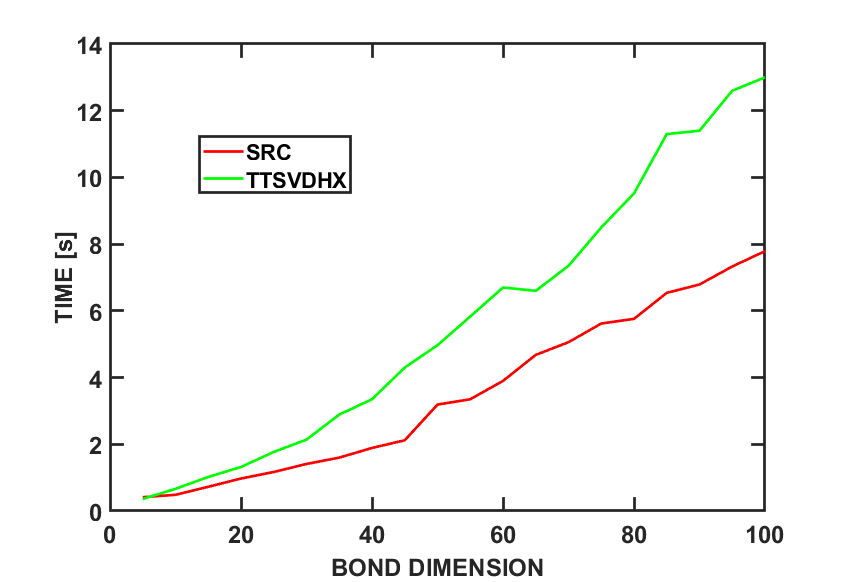}
\end{center}
\caption{Performance of TTSVD-HX and SRC in the second experiment.}
\label{grafyHX}
\end{figure}
Again, TTSVD-HX has a slightly lower approximation error, while SRC is somewhat faster.

\subsection{Chemical tensors}

\begin{figure}
\begin{center}
\includegraphics[width=0.3\linewidth]{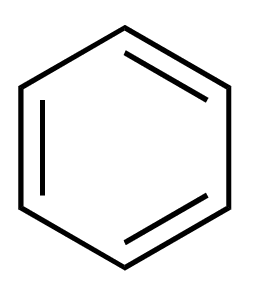}\qquad
\includegraphics[width=0.4\linewidth]{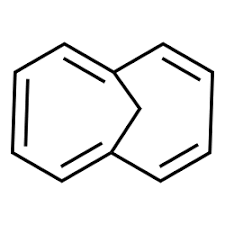}\\
\includegraphics[width=0.3\linewidth]{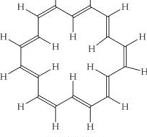}\qquad
\includegraphics[width=0.4\linewidth]{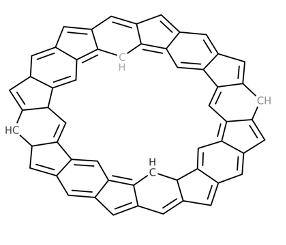}
\end{center}
\caption{Molecules under the consideration: Benzene, [10]annulene, [18]annulene, and ``azulene within azulene".}
\label{molekuly}
\end{figure}

The performance of the proposed algorithms is demonstrated on real-world data. We worked with single electron integrals $T_{ij}$ and two-electron integrals $V_{ijk\ell}$
of four cyclic aromatic molecules: benzene, [10]annulene, [18]annulene, and ``azulene within the azulene", \cite{caramori}.
The molecules are depicted in Fig. \ref{molekuly}. The number of electron sites $N=6, 10, 18$, and 56, respectively.

\begin{figure}
\begin{center}
\includegraphics[width=0.9\linewidth]{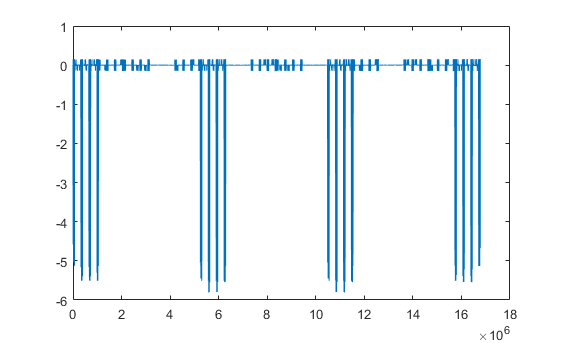}\\
\includegraphics[width=0.9\linewidth]{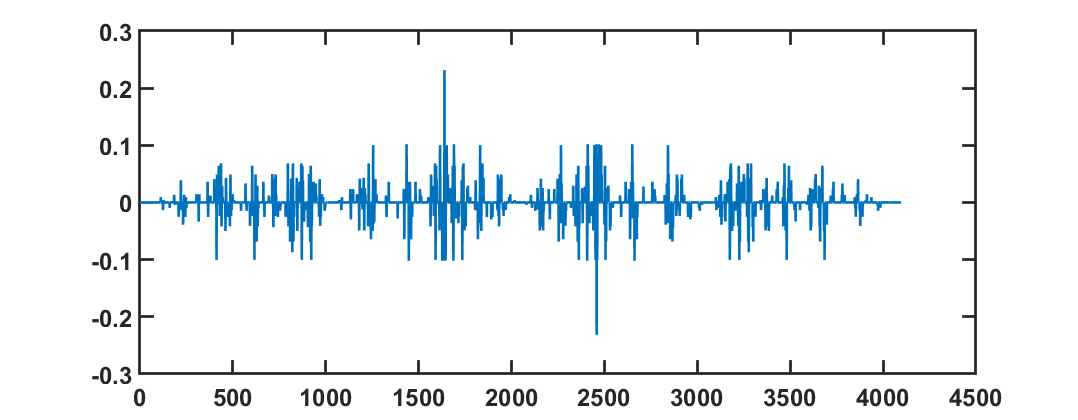}
\end{center}
\caption{Vectorized Hamiltonian, and the minimum energy wave function of benzene, respectively.}
\label{wave}
\end{figure}

The corresponding Hamiltonians, of size $16\times\ldots\times 16$ ($16^N$), are given in the CP decomposition form in (\ref{pet}). In the case of benzene ($N=6$),
we can compute the Hamiltonian in full form, because it has only $16^6=4096\times 4096$ elements. 98.9\% of the elements are null, so the tensor is sparse. The tensor in a vectorized form is shown in Fig. \ref{wave}. Both tensors are sparse, with 90.2\% and 95.0\% nulls, respectively.

The Hamiltonians are converted to the TT form using the TTSVD-CP and TTSVD-U algorithm, with aggregations. First, we divided the rank-one components into groups of size 1000,
and then the partial sums were added together. The resultant tensors are used as the initial tensors for our experiments.

Next, we used the Lanczos algorithm and the TTSVD-HX algorithm from Section IV.C to find the wave function with the minimum energy.
For now, we were able to do it only for the first three molecules. The minimum energies were confirmed by the DMRG algorithm.
Details of the experiment are summarized in Table 1.

\begin{table}
\begin{center}
\begin{tabular}{|l|c|c|c|c|c|}
\hline
molecule & N & $B_H$ & $B_x$ & TIME[s] & ENERGY\\
\hline
benzene & 6 & 87 & 64 & 0.07 & -230.7934 \\
\mbox{[10]}annulene & 10 & 112 & 450 & 22.97 & -422.4479\\
\mbox{[18]}annulene & 18 & 120 & 450 & 184.61 & -692.1739\\
\mbox{[56]}azulene & 56 & 217 & ? & ? & -2135.0183\\
\hline
\end{tabular}
\end{center}
\caption{Table 1: number of sites $N$, bond dimension $B_H$ of the Hamiltonian, bond dimension of the minimum energy wave function $B_x$, time complexity of computing the product $y=Hx$, and minimum energy of the basic state.}
\end{table}

\section{Conclusions}

    We propose a novel implementation of the original TTSVDTT algorithm for compressing tensors in TT format to a lower bond dimension, referred to as TTSVD-TT0. This method is faster than the traditional TT rounding algorithm. Second, we proposed a novel variant of the algorithm, denoted TTSVD-TT1, which is convenient when the initial bond dimension of the to-be rounded tensor are large.

    Next, we introduce three variants of the TTSVD-TT1 algorithm, each tailored to a specific scenario: (i) a tensor given as a sum of several TTs -- e.g., for converting tensors from CP format to TT format; (ii) the Hadamard product of two TTs, for which our variant is particularly efficient when the initial bond dimension is large; and (iii) a product of the form $y = Hx$, where $H$ and $x$ are both given in TT format.

    In all cases, the new algorithms perform better than or equal to the traditional method, it is faster, and its accuracy is usually better then accuracy of the randomized TT rounding methods.

\vspace*{5mm}

{\bf Acknowledgements.} The author would like to express gratitude to Dr. Libor Veis
for teaching the fundamentals of quantum chemistry and for providing the tensor
representation of the molecule involved.

The author also thanks Dr. Ivan Leonardo Perez Cabrera for his assistance with the Python implementations of the competing methods in the comparative study.


\section*{Supplemental material to \\``Algebraic Operations on Tensor Trains"\\ by Petr Tichavsky}

This material contains MATLAB implementations of the algorithms proposed in the main paper.

\begin{lstlisting}[title={{\bfseries Algorithm 1}. Scalar product of two tensors in TT format}]
function s=scalarprod(x,y)
   N=length(x); n = size(x{1},2);
   a=reshape(y{1},n,[])'*reshape(x{1},n,[]);
   for j=2:N-1
     r2 = size(x{j},1);
     [s2, n, ~] = size(y{j});
     a=reshape(a*reshape(x{j},r2,[]),s2*n,[]);
     a=reshape(y{j},s2*n,[])'*a;
   end
   s=a(:)'*reshape(y{N}*x{N}',[],1);
end
\end{lstlisting}

\begin{lstlisting}[title={{\bfseries Algorithm 2}. SVD-based TT decomposition in TT format}]
function y = TTSVDTT0(x,tol,Bmax)
y = x; I1 = 1; N=length(x);
mm=cell(1,N); xx0=1;   % mm{n}=M_n
for j=N:-1:2
    [r1,~,r2]=size(x{j});
    xx0=reshape(x{j},[],r2)*reshape(xx0,r2,[]);
    xx0=reshape(xx0,r1,[])*reshape(x{j},r1,[])';
    mm{j-1}=xx0;
end
zn=reshape(x{1},[],r1);      % z_1=x_1
for i=1:N-1
    [U,S,~]=svd(zn*chol(mm{i}),'econ');
    sval=diag(S);
    I2=min([Bmax,sum(sval>=tol*sval(1))]);
    y{i}=reshape(U(:,1:I2),I1,[],I2);
    s=size(x{i+1});
    zn=reshape(U(:,1:I2)'*zn*reshape(x{i+1},s(1),[]),[],s(end));
    I1=I2;   % zn=z_i
end
y{N} = zn;
end
\end{lstlisting}

\begin{lstlisting}[title={{\bfseries Algorithm 3}. SVD-based TT decomposition in TT format}]
function y = TTSVDTT1(x,tol,Bmax,Bmax2)
y = x; I1 = 1; N=length(x);
Sn=cell(1,N); Sn{N}=1;
for j=N:-1:2
    [r1,~,r2]=size(x{j});
    Vn=reshape(reshape(x{j},[],r2)*reshape(Sn{j},r2,[]),r1,[]);
    [U,S,~] = svd(Vn,'econ');
    sval=diag(S);
    I2=min([Bmax2,sum(sval>=tol*sval(1))]);
    Sn{j-1}=U(:,1:I2)'*S(1:I2,1:I2);
end
zn=reshape(x{1},[],r1);      % z_1=x_1
for i=1:N-1
    [U,S,~]=svd(zn*Sn{i},'econ');
    sval=diag(S);
    I2=min([Bmax,sum(sval>=tol*sval(1))]);
    y{i}=reshape(U(:,1:I2),I1,[],I2);
    s=size(x{i+1});
    zn=reshape(U(:,1:I2)'*zn*reshape(x{i+1},s(1),[]),[],s(end));
    I1=I2;   % zn=z_i
end
y{N} = zn;
end
\end{lstlisting}

\begin{lstlisting}[title={{\bfseries Algorithm 4}. SVD-based TT decomposition in SOTT format}]
function x=TTSVDU(U,tol,Bmax,Bmax2)
%
M=length(U); N=length(U{1});
ms=cell(1,N); b=cell(1,M); x=cell(1,N);
for m=1:M
    b{m}=1;
end
ms{N}=b;
for j=N:-1:2
    W=0;                % It will be matrix W_n=V_nV_n^T
    for m=1:M
        [r1,~,r2]=size(U{m}{j});
        b{m}=reshape(reshape(U{m}{j},[],r2)*reshape(b{m},r2,[]),r1,[]);
        W=W+b{m}'*b{m};
    end
    [V,S,~]=svd(W);
    B=min([Bmax2,sum(diag(S)>=tol*S(1,1))]);
    for m=1:M
        b{m}=b{m}*V(:,1:B);
    end
    ms{j-1}=b;
end
In=size(U{1}{1},2);
for m=1:M
    b{m}=reshape(U{m}{1},In,[]);
end
for i=1:N-1
    W=0; In=size(U{1}{i},2);
    for m=1:M
        W=W+b{m}*ms{i}{m};
    end
    [V,S,~]=svd(W,'econ');
    B=min([Bmax,sum(diag(S)>=tol*S(1,1))]);
    x{i}=reshape(V(:,1:B),[],In,B);
    for m=1:M
        aux=V(:,1:B)'*b{m};
        s0=size(aux,2);  s1=size(ms{i+1}{m},1);
        aux=aux*reshape(U{m}{i+1},s0,[]);
        b{m}=reshape(aux,[],s1);
    end
end
x{N}=b{1}; In=size(U{1}{N},2);
for m=2:M
    x{N}=x{N}+b{m};
end
x{N}=reshape(x{N},[],In);
end
\end{lstlisting}

\begin{lstlisting}[title={{\bfseries Algorithm 5}. SVD-based TT decomposition in CP format}]
function x=TTSVDCP(A,tol,Bmax,Bmax2)
%
% SVD-based TT decomposition of a tensor in cp format (of rank M)
% The tensor has the size KxKx...xK (K^N)
% The input factor matrices are stored in the tensor A of size KxMxN
%
[K,M,N]=size(A);
Sn=cell(1,N);
x=cell(1,N);
bb=ones(1,M);
Sn{N}=bb;
for j=N:-1:2
    bb=krb(bb,A(:,:,j));
    if size(bb,1)<size(bb,2)
       W=bb*bb';
    else
       W=bb;
    end
    [V,S,~]=svd(W,'econ');
    sval=diag(S);
    B=min([Bmax2,sum(sval>=tol*S(1,1))]);
    bb=V(:,1:B)'*bb;
    Sn{j-1}=bb;
end
In=K;
bb=A(:,:,1);
for i=1:N-1
    W=bb*Sn{i}';
    [V,S,~]=svd(W,'econ');
    sval=diag(S);
    B=min([Bmax,sum(sval>=tol*S(1,1))]);
    x{i}=reshape(V(:,1:B),[],In,B);
    bb=krb(A(:,:,i+1),V(:,1:B)'*bb);
end
x{N}=reshape(sum(bb,2),[],In);
end
\end{lstlisting}

\begin{lstlisting}[title={{\bfseries Algorithm 6}. SVD-based TT decomposition of F*G}]
function y=TTSVDHP(F,G,tol,bmax,bmax2)
%
% ttsvd algorithm for tensor given as Hadamard product of two tensors, F, G,
% given in TT format.
%
N=length(F);
Sn=cell(1,N);
Sn{N}=krb(F{N}(:,:,1),G{N}(:,:,1));
for j=N-1:-1:2
    [b1,i1,b2]=size(F{j});  % B
    [a1,~,a2]=size(G{j});   % A
    [~,c1]=size(Sn{j+1});
    C=reshape(Sn{j+1},a2,b2*c1);
    B=permute(F{j},[1,3,2]);
    A=permute(G{j},[1,3,2]);
    bux=pagemtimes(A,C);
    bux=reshape(permute(reshape(bux,a1,b2,c1*i1),[2,1,3]),b2,a1*c1,i1);
    cux=pagemtimes(B,bux);
    cux=permute(reshape(cux,b1,a1,c1*i1),[2,1,3]);
    aux=reshape(cux,a1*b1,[]);
    [U,S,~]=svd(aux,'econ');
    sval=diag(S);
    b=min([bmax,sum(sval>tol*sval(1))]);
    Sn{j}=reshape(U(:,1:b)*S(1:b,1:b),b1*a1,b);
end
[g0,g1,g2]=size(G{1});
[f0,f1,f2]=size(F{1});
zn=krb(reshape(F{1},f1,f2)',reshape(G{1},g1,g2)')';
I1=1;
for i=1:N-1
    [U,S,~]=svd(zn*Sn{i+1},'econ');
    sval=diag(S);
    I2=min([bmax,sum(sval>=tol*sval(1))]);
    y{i}=reshape(U(:,1:I2),I1,[],I2);
    [m1,m2,m3]=size(Sn{i+1});
    zn=U(:,1:I2)'*zn;
    [b1,i1,b2]=size(F{i+1});
    [a1,~,a2]=size(G{i+1});
    A=permute(G{i+1},[1,3,2]);
    B=permute(F{i+1},[1,3,2]);
    C=reshape(zn,[],b1);
    bux=pagemtimes(C,B);
    bux=permute(reshape(bux,[],a1,b2,i1),[1,3,2,4]);
    cux=pagemtimes(reshape(bux,[],a1,i1),reshape(A,a1,a2,i1)); %
    cux=reshape(cux,[],b2,a2,i1);
    zn=reshape(permute(cux,[1,4,3,2]),[],a2*b2);
    I1=I2;
end
y{N}=reshape(zn,I2,[]);
end
\end{lstlisting}

\begin{lstlisting}[title={{\bfseries Algorithm 7}. SVD-based TT decomposition of product Hx}]
function G=TTSVDHX(F,H,tol,bmax,bmax2)
%
N=length(F);
[bf,I2]=size(F{N});
[b,I]=size(H{N});
Sn=cell(1,N);
Sn{N}=permute(reshape(F{N}*reshape(H{N}',I2,[]),bf,[],b),[1,3,2]);
for j=N-1:-1:2
    [b1,i1,b2]=size(F{j});
    [b3,b34,b4]=size(H{j});
    [~,b5,b6]=size(Sn{j+1});
    aux=reshape(reshape(F{j},[],b2)*reshape(Sn{j+1},b2,[]),b1,i1,b5,b6);
    aux=reshape(permute(aux,[2,3,1,4]),i1*b5,[]);
    bux=reshape(permute(reshape(H{j},b3,i1,[],b5),[1,3,2,4]),[],i1*b5);
    aux=reshape(permute(reshape(bux*aux,b3,[],b1,b6),[3,1,2,4]),b1,b3,[]);
    b7=b34/i1*b6;
    [U,S,~]=svd(reshape(aux,[],b7),'econ');
    sval=diag(S);
    b=min([bmax,sum(sval>tol*sval(1))]);
    Sn{j}=reshape(U(:,1:b)*S(1:b,1:b),b1,b3,b);
end
G=F; I1=1; b=zeros(1,N);
F{1}=squeeze(F{1});
[f1,f3]=size(F{1});
[sh0,sh1,sh2]=size(H{1});
Hi=reshape(F{1}'*reshape(H{1},f1,[]),I1,f3,[],sh2);
for i=1:N-1
    [m1,m2,m3]=size(Sn{i+1});
    [h0,h1,h2,h3]=size(Hi);
    bux=permute(Hi,[1,3,2,4]);
    [h4,h5,h6]=size(H{i+1});
    [s45,s5,s6]=size(F{i+1});
    aux=reshape(bux,[],m1*m2)*reshape(Sn{i+1},m1*m2,m3);
    [U,S,~] = svd(aux,'econ');
    sval=diag(S);
    I2=min([I2,sum(sval>=tol*sval(1))]);
    G{i}=reshape(U(:,1:I2),I1,h2,I2);
    hu=size(U,1);
    aux=reshape(U(:,1:I2)'*reshape(bux,hu,[]),[],h4);
    bux=reshape(permute(reshape(aux,I2,h1,h4),[1,3,2]),I2*h4,h1)*reshape(F{i+1},h1,[]);
    Hi=reshape(permute(reshape(bux,I2,h4*s5,s6),[1,3,2]),I2*s6,h4*s5)*reshape(H{i+1},h4*s5,[]);
    Hi=reshape(Hi,I2,s6,[],h6);
    I1=I2;
end
G{N}=reshape(Hi,I2,[]);
end

\end{lstlisting}

\begin{lstlisting}[title={{\bfseries Algorithm 8}. Khatri-Rao product}]
function AB = krb(A,B)
%
[I,F]=size(A);
[J,F1]=size(B);
AB=zeros(I*J,F);
for f=1:F
   ab=B(:,f)*A(:,f).';
   AB(:,f)=ab(:);
end
end
\end{lstlisting}

\end{document}